# THE PHYSICS OF MOTORCYCLES AND FAST BICYCLES : LEAN, STABILITY AND COUNTER-STEERING

## B Shayak

Department of Theoretical and Applied Mechanics,
School of Mechanical and Aerospace Engineering,
Cornell University,
Ithaca – 14853,
New York State, USA

sb2344@cornell.edu , shayak.2015@iitkalumni.org

\*

## Abstract

In this work, I will obtain the system of nonlinear equations that correctly describes the motion of motorcycles and fast bicycles (above 30 km/hr) for the first time in literature. I will use it to calculate the lean angle during a turn and prove that the motion of the vehicle is unconditionally stable at all operating speeds under consideration. I will then employ it to give a quantitative model of counter-steering – the phenomenon by which a turning mobike first goes the wrong way and then starts going the right way.

\*   \*   \*   \*   \*

## Introduction

The dynamics of bicycles and motorcycles have been of interest for more than a hundred years. Among the first to analyse this motion was FRANK WHIPPLE [1]. His calculations for a bicycle were somewhat simplified and clarified in the treatise by FELIX KLEIN and ARNOLD SOMMERFELD [2]. In this seminal work, the authors use linearized equations of motion (EOM), perturbing off a state where the bike is running straight; they conclude that bike is stable only in the speed range 16 to 20 km/hr and unstable everywhere else. The high speed instability is in the lean angle; according to this work, a small perturbation in this angle grows in time and causes the bike to capsize. SOMMERFELD admits that this instability "is not observed in practice". He also concludes that gyroscopic effects play a small but significant role in the stabilization of the bicycle. A recent series of works by JIM PAPADOPOULOS, ANDY RUINA, AREND SCHWAB [3,4] and others however has shown SOMMERFELD to be wrong. These later authors have proved that in fact a bicycle can be stable in the absence of gyroscopic effects. The primary factors affecting stability are the inclination of the steering axis from the vertical and the distribution of the masses of the frame, the driver and the handlebars. To prove these claims, these authors again use linearized equations of motion, working in the regime where the bike's motion is straight, and the lean and steering angles shallow. To quote them, "all derivations to date, including this one, involve ad hoc linearization as opposed to linearization of full nonlinear equations. No one has linearized the full implicit nonlinear equations (implicit because there is no reasonably simple closed form expression for the closed kinematic chain) into an explicit analytical form by either hand or computer algebra". It is noteworthy that this analysis retains the unphysical upper bound on speed of stable operation found in [1,2].

Parallelly, let us examine the motorcycle literature. The pioneering paper here was by ROBIN SHARP [5], who used a Lagrangian procedure. Like SOMMERFELD, he has derived linear equations describing the stability of the mobike moving in a straight line. These equations transition from unstable to stable and then back to unstable as the speed increases. Again like the bike works, the high-speed instability lies in the lean angle; SHARP's mobike is stable in the speed range 20 to 40 km/hr and unstable everywhere else. His original paper has not treated the behaviour on turns. A turning motorcycle model has been considered by C KOENEN and H B PACEJKA [6] who use 28 state variables to characterize the mobike's motion. Linearizing about an ad hoc turning fixed point, they obtain an EOM which takes 21 pages to print. The mobike following this EOM remains stable in the speed range 30 to 65 km/hr. Several authors [7-11] have also attempted a nonlinear description of the mobike but have not succeeded in actually writing an EOM which is free of unknown forces and torques. These implicit EOM's can apparently be generated and solved 'in real time' on a sufficiently advanced computer. Their linearization again yields an instability in the lean angle, at speeds similar to the works I mentioned before.

This brief literature review (a large collection of review material can be found in [12] but the principal findings are all included in the above summary) raises several pressing questions :

(*a*) Why is the model bike/mobike unstable at typical operating speeds ? Racing mobikes reach 150-200 km/hr on turns and 300+ on straights, while racing bicycles can go upto 70-80 km/hr or more – 4 to 5 times of the stable MPS according to the literature. The authors of such models claim that the instability is 'slight' because the offending eigenvalue is small and that an imperceptible amount of active control by the driver is required to ensure stability. The rock solid stability of a racing mobike defies the control claim. Dynamical instability would imply that a single error by a driver during a race would result in a deadly crash. The more so in turns where one or two extra degrees of lean would be sufficient for collision with ground, and a catastrophe. The claim of 'weak' instability is also unsatisfactory – a system is completely unstable even if one eigenvalue is positive, however small it be. In most dynamical systems, the transition from stability to instability occurs through a sign change of a single eigenvalue or pair of eigenvalues and not a migration of eigenvalues en masse across the origin or imaginary axis. In the Lorenz model [13], the transition from stable fixed point to chaos occurs through one pair of eigenvalues just breaching the imaginary axis in a Hopf bifurcation while the third eigenvalue remains negative. In coupled oscillator systems [14] the stability of a synchronized state is evaluated by considering only the largest Lyapunov exponent of the system; destabilization occurs the moment it crosses the value zero. In delay equations [15] there are an infinite number of eigenvalues – the transition from stability to instability occurs when the first pair crosses the imaginary axis from negative to positive, while all other eigenvalues remain negative. The motorcycle cannot be a counterexample to this most basic dynamical principle.

(*b*) The high speed instability also amounts to a conflict with the results for a coin rolling on a table, where the system is found to become increasingly stable as the speed of rolling increases. The coin is a very much related problem to the mobike – each of the two wheels of a mobike is like a rolling coin. The contrary motions of the coin and the mobike are surprising to say the least.

(*c*) How can we calculate the equilibrium lean angle during the turning of the motorcycle ? An equilibrium state usually comes about as a fixed point of a nonlinear system – where there is no nonlinear equation, how can there be a fixed point ? A candidate lean angle formula exists (it will come up in Section 2) but its derivation is less than credible.

(*d*) What is the dynamics of transition from a straight running state to a turning state and vice versa ? Such a transition is an inherently nonlinear phenomenon as it involves a sweep of the lean angle over a large range. A linearized theory about any one particular angle cannot hope to explain this transitional dynamics.

In our hunt for the nonlinear dynamics of the bike/mobike and our answer to the above queries, we will go to a regime opposite to the one considered by the bicycle researchers [1-4]. We will focus exclusively on the motorcycle, where the operating speeds are vastly higher and the wheels also larger. Then, we will work in the high-speed regime where the gyroscopic action is dominant, and other effects assume appropriately humble roles. The objective of this Article is thus to write a nonlinear equation of motion of a motorcycle in the gyroscopy-dominated limit. Using it I will calculate the lean angle, find the stability (correctly), and obtain the dynamics of turn entry which includes a paradoxical phenomenon called counter steering.

The outline of this lengthy Article is as follows. In Section 1 I will solve the auxiliary problem of a coin rolling in a circle on a table. This will act as an introduction to many of the concepts which will appear again in the main analysis, with greatly increased difficulty. In Section 2 I will set up the geometry of the motorcycle and explicitly solve the constraints. This will undoubtedly be the single biggest step of the whole derivation – the Literature's inability to solve the closed kinematic chain is essentially a failure to obtain the constraints in closed form. In Section 3 I will write the force balance equations and in Section 4 the torque balance equations. Finally, in Section 5 I will combine the results of the preceding Sections to present a unified nonlinear EOM of the motorcycle, and compare the predictions of this EOM with reality.

## 1  The circular motion of a rolling coin

It is a common fact that if a coin is released with some care on a flat table, it goes into a state where it leans inwards while describing a broad circular trajectory on the table (Fig. 1). A simple calculation of the lean angle has been prescribed as an exercise in the excellent textbook by DANIEL KLEPPNER and ROBERT KOLENKOW [16]; in this Section I will also prove that the motion of the coin is stable.

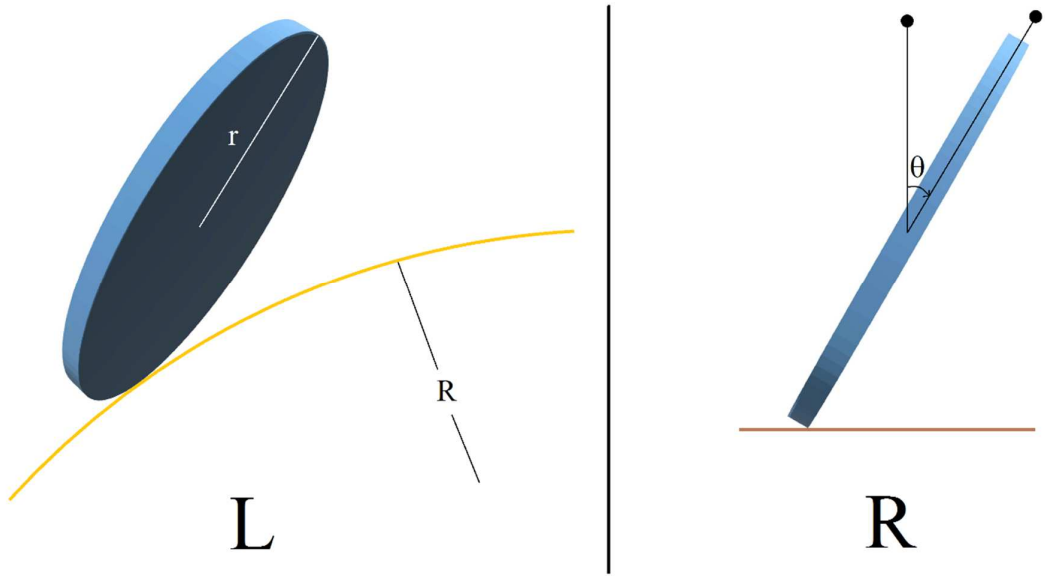

Figure 1 : *Panel L shows the 3D view of the coin rolling on the table. The yellow curve is the circular trajectory on the table, and is traversed clockwise. Panel R shows the back view (orthographic). The lean angle* θ *can be seen here.*

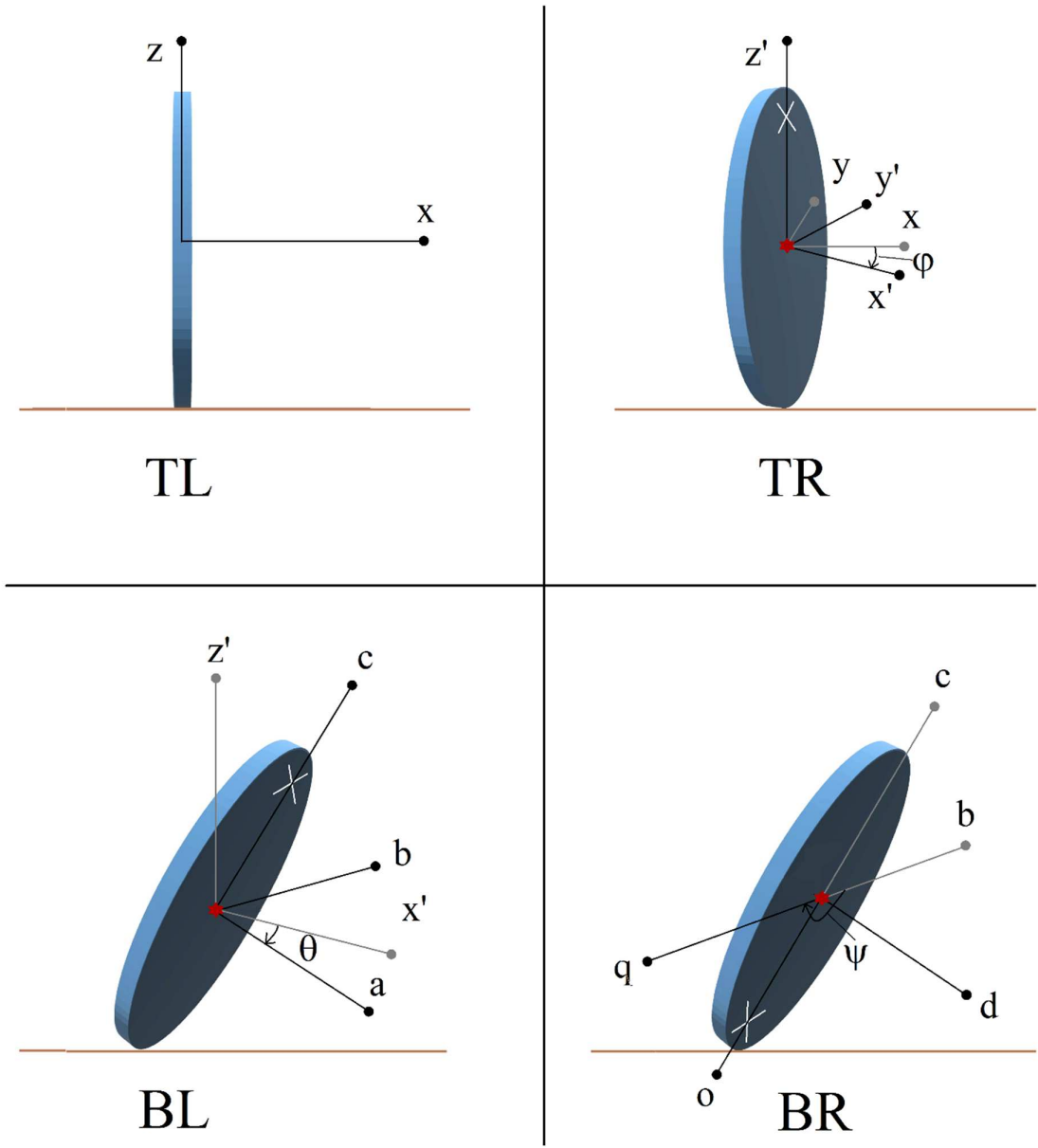

Figure 2 : *The successive Eulerian rotations which transform from the basis* d,q,o. *TL shows the reference configuration – z is along the vertical,* y *along the coin's motion and x points to starboard. The first rotation is yaw through φ about z-axis, shown in TR. This produces the basis* x',y',z' *– the old axes can be seen in grey and the new ones in black. The z and z' axes of course coincide. A cross has been marked on the coin to visually break its circular symmetry and make its orientation apparent. BL shows the second rotation – bank through θ about y'-axis to produce* a,b,c. *Finally, BR shows the last rotation, spin ψ about the* a-axis *to produce* d,q,o.

We consider a coin of radius *r* and mass *m* rolling with translational speed *V* on a circle of radius *R*. There is no way we can obtain stability of motion from diagrams and semi-quantitative arguments which Kleppner expects the readers of his book (first or second year undergrads) to use. Hence we now invoke the full machinery of 3D rotations. The greatest subtlety lies in the choice of origin and axis conventions. Let us do the axes first, assuming for now that the origin is at the CM. The coin clearly has three types of angular motions – the slow precession like thing as it changes direction during circular motion, the obvious fast spin about its axis and the lean which if unchecked will cause it to topple. These are about more or less perpendicular axes as in Fig. 1 – precession about the vertical, spin about something close to a horizontal in the plane of the page and lean about an axis out of the page plane. This suggests an Euler angle convention where the three rotations are about three different axes, called a BRIAN TAIT convention (we note that a different convention has been used by FRANK MOON [17] to analyse this problem). Further, since the coin is symmetric, we want to avail of the version of Euler's equations which arises from evaluation of the material derivative of angular momentum in a frame which does not possess the rotation about the axis of symmetry of the body. Accordingly, the last of the rotations in the Eulerian chain must be the spin or pitch about the symmetry axis. (As per the Tait convention jargon, 'precession' becomes 'yaw', 'lean' becomes 'bank' and the rolling of the coin becomes 'pitch'.) Hence, starting from a lab-fixed $x,y,z$ basis with $z$ along the vertical, $y$ along the coin's motion and $x$ pointing directly to starboard, we first yaw through $\varphi$ about $z$-axis to get the basis $x',y',z'$. Then we bank about $y'$-axis through $\theta$ to form $a,b,c$. Lastly we pitch about $a$-axis through $\psi$ to form $d,q,o$. This describes a $Z,Y,X$ rotation sequence, shown in Fig. 2.

The next issue is that of the origin. Although in the above discussion I assumed that the origin was the centre of mass (CM), I could just as well translate it to a different point. A tempting choice is the contact point between the coin and the ground – ($a$) its $z$-coordinate is fixed from the constraints, and ($b$) since all the ground forces pass through there, we will not need to worry about calculating the normal reaction or the friction as the CM wiggles about. Although this origin is accelerated, that effect can be factored in through an extra non-inertial force at the CM of the body. However this origin is unsuitable because specifying $\varphi,\theta,\psi$ about the Eulerian axes and that origin will NOT take us to any desired point on the coin's actual trajectory. The first two will go through smoothly but $\psi$ will be a problem. The default state (all angles zero) in Fig. 2 means that the coin is on the table and the cross engraved on it is at the top. Now consider $\varphi=\theta=0$ and $\psi=180°$. The configuration we want this to describe is one where the coin is on the table and the cross is at the bottom, as in Panel BR, minus the yaw and bank. However, these angles about this origin will take us to a configuration where the cross is at the bottom and the coin is under the table. To get around this problem, we must keep the origin as the CM and calculate its acceleration explicitly.

In the circular motion state, the centripetal acceleration of the CM is $\left( V^2 / r \right) \hat{\mathbf{x}}'$, this unit vector being the one which is in the horizontal plane but directed normal inwards to the coin's instantaneous motion. Extra accelerations occur on account of the banking. If the CM is held fixed and the bank applied, then the bottom of the coin will no longer touch the table i.e. the problem constraints will get violated. Hence, banking must be accompanied by translation of the CM in such a manner that the bottom of the coin remains on the table. We note that this constraint violation does not occur due to the yaw and pitch motions. Recognizing that the bank angle is the same irrespective of whether the origin is at the CM or at the contact point on the ground, for this sub-calculation we shift to the latter. About this point, the instantaneous velocity of the CM on account of the banking is $\mathbf{v}_{\text{bank}} = r\dot{\theta}\hat{\mathbf{a}}$, but this formula is valid only in the rotating dual of the *a,b,c* frame. To evaluate acceleration in the stationary *a,b,c* frame we need to use the material derivative, $d\mathbf{v}_{\text{bank}} / dt = \partial \mathbf{v}_{\text{bank}} / \partial t + \boldsymbol{\omega}^f \times \mathbf{v}_{\text{bank}}$, where $\boldsymbol{\omega}^f$ is the angular velocity of the dual frame.

To calculate that, we first obtain the *a,b,c* representation of the coin's total angular velocity $\boldsymbol{\omega}$. As mentioned above, the *a,b,c* frame follows from *x,y,z* after rotation through $\varphi$ about *z* then $\theta$ about *y'*. Since $\dot{\varphi}$ is about the *z*-axis or *z'*-axis, it can be projected into *a,b,c* as

$$\begin{bmatrix} \omega_{\text{yaw},a} \\ \omega_{\text{yaw},b} \\ \omega_{\text{yaw},c} \end{bmatrix} = \mathbf{Y}(\theta) \begin{bmatrix} 0 \\ 0 \\ \dot{\varphi} \end{bmatrix} \quad . \tag{1}$$

The banking is about *y'*-axis which is *b*-axis while the coin's spin or pitch is about *a*-axis hence these do not require any axis transformations. Putting these together,

$$\omega_a = \dot{\psi} - \dot{\varphi}\sin\theta \quad , \tag{2a}$$

$$\omega_b = \dot{\theta} \quad , \tag{2b}$$

$$\omega_c = \dot{\varphi}\cos\theta \quad . \tag{2c}$$

Recall that the *a,b,c* frame shares the yaw and bank of the coin but not its spin, which is about *a*-axis. Hence $\boldsymbol{\omega}^f$ is

$$\omega_a^f = -\dot{\varphi}\sin\theta \quad , \tag{3a}$$

$$\omega_b^f = \dot{\theta} \quad , \tag{3b}$$

$$\omega_c^f = \dot{\varphi}\cos\theta \quad . \tag{3c}$$

Then,

$$\begin{aligned} \frac{d\mathbf{v}_{\text{bank}}}{dt} &= \dot{\mathbf{v}}_{\text{bank}} + \boldsymbol{\omega}^f \times \mathbf{v}_{\text{bank}} \\ &= r\ddot{\theta}\hat{\mathbf{a}} + \left( (...)\hat{\mathbf{a}} + \dot{\theta}\hat{\mathbf{b}} + \dot{\varphi}\cos\theta\hat{\mathbf{c}} \right) \times r\dot{\theta}\hat{\mathbf{a}} \quad . \\ &= r\ddot{\theta}\hat{\mathbf{a}} + r\dot{\varphi}\dot{\theta}\cos\theta\hat{\mathbf{b}} - r\dot{\theta}^2\hat{\mathbf{c}} \end{aligned} \tag{4}$$

This is the acceleration of the CM of the coin on account of the banking motions.

The CM's total acceleration (centripetal plus banking) will be caused by the resultant of gravity and the force coming from the ground; in the latter term, if we already include the default term $+mg\hat{\mathbf{z}}$, then all the additional force causing the acceleration can also be attributed to the ground. Hence we obtain the resultant force on the coin from the ground by adding the contributions coming from weight balance, centripetal acceleration and acceleration due to banking (G denotes ground) :

$$\mathbf{F}_G = \frac{mV^2}{R}\hat{\mathbf{x}}' + mg\hat{\mathbf{z}} + mr\left(\ddot{\theta}\hat{\mathbf{a}} + \dot{\varphi}\dot{\theta}\cos\theta\,\hat{\mathbf{b}} - \dot{\theta}^2\hat{\mathbf{c}}\right) \quad . \tag{5}$$

The torque of this force about the CM can be calculated easily.

Now taking the material derivative of $\mathbf{L}$ in the $a,b,c$ frame yields

$$I_a\dot{\omega}_a = T_a \quad , \tag{6a}$$

$$I\dot{\omega}_b + I_a\omega_a\omega_c^f - I\omega_c\omega_a^f = T_b \quad , \tag{6b}$$

$$I_c\dot{\omega}_c + I\omega_b\omega_a^f - I_a\omega_a\omega_b^f = T_c \quad . \tag{6c}$$

We note that $I_a = mr^2/2$ and $I = mr^2/4$.

The last step is calculating the torques $T_a$, $T_b$ and $T_c$ i.e. the right hand side (RHS) of the EOM. The only torque is exerted by the force coming from the ground, which we evaluated in (5). Since $z$-axis is the same as $z'$-axis, $mg$ can be written to be about that latter axis. Then, gravity and centripetal acceleration can be projected into $a,b,c$ with a single application of the matrix $\mathbf{Y}(\theta)$. The acceleration due to banking is already in that basis so we can leave it untouched. Thus we have

$$\begin{bmatrix} F_{Ga} \\ F_{Gb} \\ F_{Gc} \end{bmatrix} = \mathbf{Y}(\theta)\begin{bmatrix} mV^2/R \\ 0 \\ mg \end{bmatrix} + mr\begin{bmatrix} \ddot{\theta} \\ \dot{\varphi}\dot{\theta}\cos\theta \\ -\dot{\theta}^2 \end{bmatrix} \quad , \text{ or} \tag{7a}$$

$$\begin{bmatrix} F_{Ga} \\ F_{Gb} \\ F_{Gc} \end{bmatrix} = \begin{bmatrix} -mg\sin\theta + \dfrac{mV^2}{R}\cos\theta + mr\ddot{\theta} \\[2mm] mr\dot{\varphi}\dot{\theta}\cos\theta \\[2mm] mg\cos\theta + \dfrac{mV^2}{R}\sin\theta - mr\dot{\theta}^2 \end{bmatrix} \quad . \tag{7b}$$

This must be crossed with the position vector from the CM to the contact point with the ground, which is $-r\hat{\mathbf{c}}$. Then,

$$T_a = mr^2\dot{\varphi}\dot{\theta}\cos\theta \quad , \tag{8a}$$

$$T_b = mgr\sin\theta - \frac{mV^2r}{R}\cos\theta - mr^2\ddot{\theta} \quad , \tag{8b}$$

$$T_c = 0 \quad . \tag{8c}$$

With this step, all the ingradients of the EOM have been determined.

Substituting $\boldsymbol{\omega}$, $\boldsymbol{\omega}^f$ and $\mathbf{T}$ into (6) leads to the overall EOM for the rolling coin, where $\Omega = \dot{\varphi}$ denotes the frequency of yaw :

$$I_a\left(\ddot{\psi} - \dot{\Omega}\sin\theta - \Omega\dot{\theta}\cos\theta\right) = mr^2\Omega\dot{\theta}\cos\theta \quad , \tag{9a}$$

$$I\ddot{\theta} + I_a\left(\dot{\psi} - \Omega\sin\theta\right)\Omega\cos\theta + I\Omega^2\cos\theta\sin\theta = mgr\sin\theta - \frac{mV^2r}{R}\cos\theta - mr^2\ddot{\theta}, \tag{9b}$$

$$I\left(\dot{\Omega}\cos\theta - \Omega\dot{\theta}\sin\theta\right) - I\dot{\theta}\dot{\Omega}\sin\theta - I_a\left(\dot{\psi} - \Omega\sin\theta\right)\dot{\theta} = 0 \quad . \tag{9c}$$

We are interested in a fixed point where $\dot{\psi}$, $\theta$ and $\Omega$ are constant; clearly, this fixed point satisfies

$$I_a\left(\dot{\psi}^* - \Omega^*\sin\theta^*\right)\Omega^*\cos\theta^* + I\Omega^{*2}\cos\theta^*\sin\theta^* = mgr\sin\theta^* - \frac{mV^2r}{R}\cos\theta^* \quad . \tag{10}$$

Under the assumption of fast spin i.e. $\dot{\psi}^* >> \Omega^*$, and using $\dot{\psi}^* = V/r$ and $\Omega^* = V/R$, we recover Kleppner's answer of $\theta^* = \arctan\left(3V^2/2gR\right)$, an important consistency check. Note that the rates of yaw and pitch are both negative for the situation shown in Fig. 1 so the signs here are correct.

Applying the fast spin approximation, we rewrite (9). Under this assumption, $\omega_a$ becomes equal to $\dot{\psi}$. From (9a), its derivative is a product of two small terms, hence at the largest order (which is what we care about) it is a constant of the motion; since it is negative, we call it $-v$. This leads to a simplification of the EOM. Further, we need to eliminate $V$ in the RHS of (9b); since $\Omega = V/R$, $V^2r/R = \Omega^2rR$. This leads to the simplified system

$$I\ddot{\theta} - I_a v\Omega\cos\theta - mgr\sin\theta + mrR\Omega^2\cos\theta + mr^2\ddot{\theta} = 0 \quad , \tag{11a}$$

$$I\left(\dot{\Omega}\cos\theta - \Omega\dot{\theta}\sin\theta\right) + I_a v\dot{\theta} = 0 \quad . \tag{11b}$$

Since $v = V/r$ and $\Omega = V/R$, the fast top assumption is good if $R >> r$. In a typical scenario, the radius of the motion is about ten times that of the coin itself, and this approximation should work fine.

Linearizing (11) we get, where $Y = \dot{\theta}$,

$$\left(I + mr^2\right)\ddot{Y}$$
$$= \left[\begin{array}{l} mgr\cos\theta^* - I_a v\Omega^*\sin\theta^* + mrR\Omega^{*2}\sin\theta^* \\ +\dfrac{1}{I\cos\theta^*}\left(I\Omega^*\sin\theta^* - I_a v\right)\left(I_a v\cos\theta^* - 2mrR\Omega^*\cos\theta^*\right) \end{array}\right]Y \quad . \tag{12}$$

The motion will be stable if the above describes a harmonic oscillator, and unstable if it is a harmonic repeller. Clearly, the coefficient on the LHS is positive, but the RHS has a profusion of terms of varying signs. In the fast coin regime, some terms will be larger than others, with $v$ being the driving factor. Before comparing terms, we note that $\Omega^*R = -vr$ from the rolling without slipping condition – this enables us to express everything in terms of $v$ and $r$. Using this, the largest term in the RHS is $-I_a\left(I_a + 2mr^2\right)v^2/I$, which is negative implying that (12) describes a harmonic oscillator and not a repeller. Hence the circular motion of the coin is stable, in agreement with experimental observations and with literature [17]. We note that the 'stiffness' of the 'spring' in (12) becomes larger and larger as its speed increases. This is a very plausible conclusion – since the rotation is providing the stability, faster rotation implies greater stability.

## 2 Motorcycle geometry and the equations of constraint

Before starting the analysis proper let me mention the oft-quoted formula for the lean angle which appears in many websites as well as technical articles. This formula is

$$\theta = \arctan \frac{V^2}{gR} \quad , \tag{xx}$$

where $V$ is the forward speed of the mobike, $R$ the radius of the turn and $g$ the acceleration due to gravity. Equation (xx), whose number highlights that it is not part of the main analysis, is obtained from a 2D torque balance, treating the mobike as a stick object pivoted to the ground. The pivot rotates during the turn and is hence accelerated, and an easy balance between the torques of gravity and centrifugal force shown in Fig. 3 leads to the result. If we look closer we can see what will happen when the stick mobike is perturbed slightly from its equilibrium position. If it is displaced downward, then the torque of gravity will increase, that of centrifugal force will decrease and the mobike will go further downwards. This is not what real mobikes do and so this stick model does not seem to make too much sense. Nevertheless I will refrain from commenting the veracity of the formula (xx) itself until my own equations of motion are out.

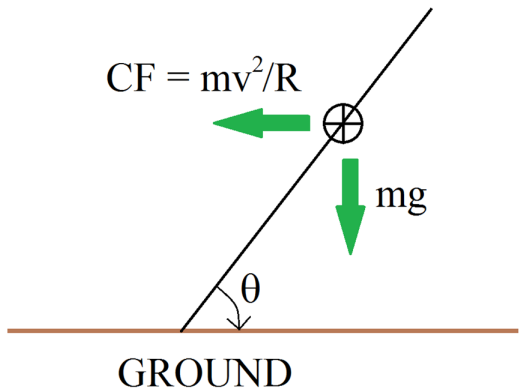

Figure 3 : *A 2D model of a motorcycle. The view is from the back and the turn is to the starboard. The torque balance is attempted about the wheel-ground contact point; since this is moving, there is a non-inertial centrifugal force in this frame, acting at the CM. There is also gravity. The torque of gravity is* mg*sin*θ *while that of centrifugal force is (*mV²/R)*cos*θ*; the two are in opposite directions and balance when tan*θ=V²/gR.

Although the primary technique in the literature to analyse motorcycle dynamics has been to use Lagrange's equations as the starting point (the only exception is SOMMERFELD who goes Newtonian), I will not adopt that approach here. Firstly, the constraints involved in this problem are non-holonomic [18,19] – they are a combination of the Chaplygin skate and the stationary contact point constraint – and this makes the Lagrangian formulation extremely difficult. But in a Newtonian formalism, the constraints can be accounted for naturally with no special manoeuvres and machinations. Secondly, a direct force and torque balance conveys enormous insight into the behaviour of the system, and this qualitative feel for the mobike's motions will guide the mathematical development along a path of least resistance towards the final solution.

I will model the mobike as shown in Fig. 4 and consider a leaning turn to the starboard. This paper mobike is composed of three rigid bodies : the rear wheel, the front wheel and the frame which connects the two wheels together and accommodates the driver. Let body B1 be the rear wheel, B2 the front wheel and B3 the frame. I will assume that both the wheels are 'ideal' i.e. they are disks of negligible thickness, are pivoted at dead centre, have one principal axis along the axis of geometrical symmetry and the other two in the plane, and have their in-plane moments of inertia equal. I will model the frame as a massless 'T'-shaped truss with the long arm running along the mobike's long axis from the centre of B1 to that of B2. We will label these centres as $O_1$ and $O_2$. At the centre of the long arm there is a point mass corresponding to engine, fuel and the 'bulk of the driver'. The short arm branches out from this point, horizontal in the unbanked state and perpendicular to the mobike's axis and carries a mass at its end. This mass accounts for the possibility of the driver's leaning out of his mobike during the turn. During the analysis, I will discuss the implications of some of these approximations, and the means to circumvent them.

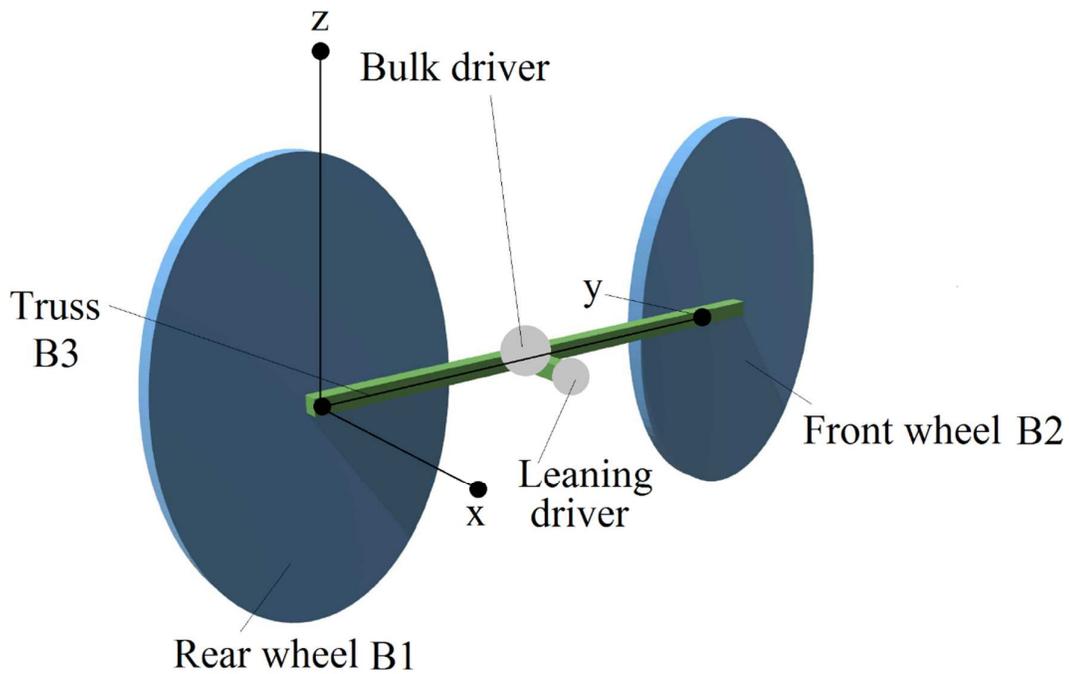

Figure 4 : *The mobike in its reference configuration where all Euler angles are zero. The alignment of the reference* x,y,z *basis can be seen. The description of the various components has been given in the bulk text.*

We let the various objects have the following parameters :

- Rear wheel mass : $m_1$
- Rear wheel radius : $r_1$
- Rear wheel moment of inertia about the axis of symmetry : $I_{1a}$ ('$a$' for 'axial')

- Rear wheel moment of inertia about a perpendicular axis : $I_{1s}$ ('$s$' for 'symmetric')
- Front wheel mass : $m_2$
- Front wheel radius : $r_2$
- Front wheel moment of inertia about the axis of symmetry : $I_{2a}$
- Front wheel moment of inertia about a perpendicular axis : $I_{2s}$
- Truss long arm length : $2l_3$. This is called 'wheelbase' in the literature.
- Truss short arm length : $l_4$
- Truss total mass : $m_3$
- Leaning driver particle mass : $m_4$
- Mass ratio : $\eta = m_4/m_3$
- Truss moment of inertia about $a_3$-axis (to be defined in a moment) : $I_{3a}$
- Truss moment of inertia about $b_3$-axis : $I_{3b}$
- Truss moment of inertia about $c_3$-axis : $I_{3c}$

As in the coin, the orientation of each of the three constituent bodies can be characterized using the Tait convention angles $\varphi$, $\theta$ and $\psi$, defined in the same manner and chosen in the same order. Since there are three bodies in the mobike, each of the angles here will acquire a subscript corresponding to the body number.

Apart from the subscript '1's, the reference configuration and Euler angles for the rear wheel will be identical to those of the coin. The truss also poses no problem – zero rotation is the reference configuration of the mobike (with the long arm along $y$-axis) and then yaw, bank and pitch. But the front wheel is a source of difficulty. Since in a real mobike its steering axis is inclined (has a combination of $y$ and $z$ components), that must be chosen as one of the cardinal axes in the reference configuration. Accordingly, a new reference basis $x_2,y_2,z_2$ must be defined for this wheel, then the Euler angles and the $a_2,b_2,c_2$ basis defined appropriately, and extra rotation matrices used to go from $x_2,y_2,z_2$ to $x_1,y_1,z_1$. To avoid this complication, I will take advantage of the fact that we are working with a fast mobike here. The bicycle studies have clearly shown that the effect of this inclination ("caster trail") is to generate stability at low speeds. At high speeds, when the contraption should be inherently stable from gyroscopy, this effect will play a secondary role. Accordingly I will assume that the steering axis of the front wheel is vertical. This will enable all three bodies to acquire the same reference basis.

After the reference basis issue is settled, let me turn to the question of how many angles there really are. Although there are nine angle variables $\varphi$, $\theta$, and $\psi$ for each of B1, B2 and B3, many of them are in fact not independent. From practical observation, the rear wheel is mounted rigidly to the frame so far as yaw is concerned – the wheel does not have the freedom to yaw about relative to the frame. Hence

$$\varphi_1 = \varphi_3 \quad . \tag{13}$$

Because of the steering degree of freedom, the front wheel can yaw relative to the frame so $\varphi_2$ does not necessary equal these two. We let

$$\varphi_2 = \varphi_1 + \Phi \quad , \tag{14}$$

which acts as the definition of $\Phi$ (a variable of critical importance in what follows). From the geometry, $\Phi$ is actually negative if the turn is to the starboard. Now we turn to the lean. From practical observation, the mobike banks as a whole – the bank angle of the rear wheel, the front wheel and the driver are all the same (at least to very good accuracy). Hence

$$\theta_1 = \theta_2 = \theta_3 = \theta \quad . \tag{15}$$

Now suppose that the radii of the rear and front wheels are equal, $r_1 = r_2 = r$. Then the above equation implies that $O_1$ and $O_2$ are always at the same height above ground. Hence, both ends of the long arm of the truss are at the same height and this arm thus lies in the horizontal plane. This implies that the truss cannot pitch during the mobike's motions. Hence

$$\psi_3 = 0 \quad . \tag{16}$$

If the two wheel radii were unequal then every value of $\theta$ would correspond to some unique value of $\psi_3$ and although the simple form of (16) would be lost, a constraint equation would remain anyway. Equations (13-16) imply that the $a,b,c$ bases of the rear wheel and the truss are identical i.e.

$$\begin{aligned}
\hat{\mathbf{a}}_1 &= \hat{\mathbf{a}}_3 \\
\hat{\mathbf{b}}_1 &= \hat{\mathbf{b}}_3 \quad . \\
\hat{\mathbf{c}}_1 &= \hat{\mathbf{c}}_3
\end{aligned} \tag{17}$$

Since $\psi_3 = 0$, $a,b,c$ is the full body frame of the truss; since the rear wheel is symmetric, $a,b,c$ is also the basis which is most relevant for its analysis. Hence (17) is good; it will be very useful later. But where are the origins of these various bases ?

The origins of the two wheels are of course their CMs but the truss is a different story. Since it is a weirdly shaped accelerating body, conventional wisdom would have us choose its CM as the origin of rotational motions. The top view of the mobike in Fig. 5 however shows that if the back wheel is to roll without slipping, then this choice of origin contradicts the constraint (13). Suppose that the extra mass $m_4$ is zero, whereby the truss CM is at the centre of the long arm. Suppose also that the bank angle is zero. Then, if the truss has a yaw angular velocity about the CM, the centre of the rear wheel will acquire a velocity component along the $x'_1$-axis. Since from (13), the orientation of the truss and the rear wheel are parallel, the pitch (spin) of the wheel will impart to all points on its rim an additional velocity which lies only in the $y'_1, z'_1$ plane. Rolling without slipping means that the combined velocity of one of these points (the one in contact with the ground) must be zero. However, the two contributions from yaw and pitch are in perpendicular directions so they cannot have a vanishing resultant anywhere on the wheel. Hence rolling without slipping amounts to a contradiction with the constraint (13).

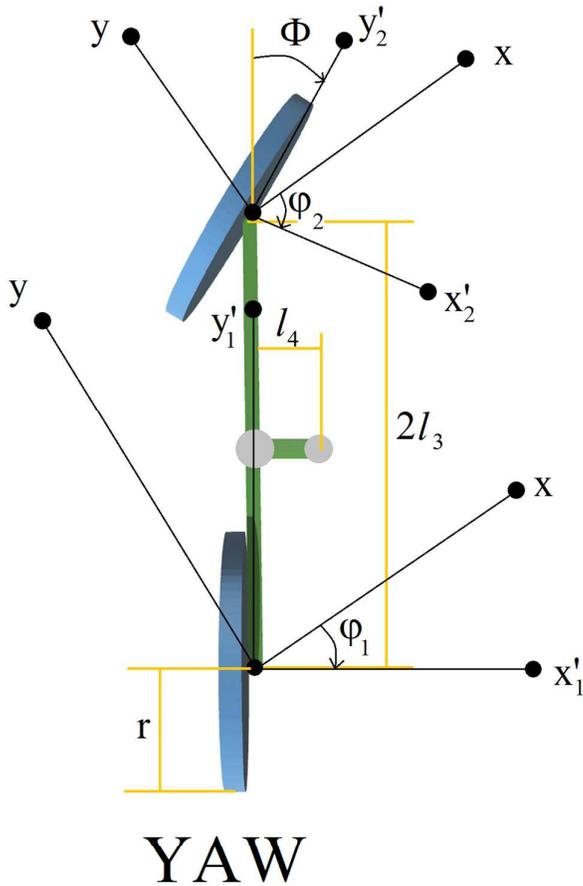

**YAW**

Figure 5 : *The rotations φ and θ. The upper panel shows the mobike in top view, to bring out the yaw. Recall that the yaw is applied on the reference configuration. For schematic clarity, the reference axes have been tilted and the bike long axis is parallel to the page axis. The z-axis points out of the page. The relations (13), (14) and (17) are apparent from this view. The definitions of $l_3$ and $l_4$ are also clear. Note that the difference $\Phi=\varphi_2-\varphi_1$ has been grossly exaggerated in this Figure – its actual value is of the order of $1^o$ ! The lower panel shows the back view to indicate the bank which is applied to the yawed configuration. I have labelled only the $a_1,b_1,c_1$ basis for clarity; the 2-basis should be apparent from pattern recognition. The $b_1$-axis points into the page.*

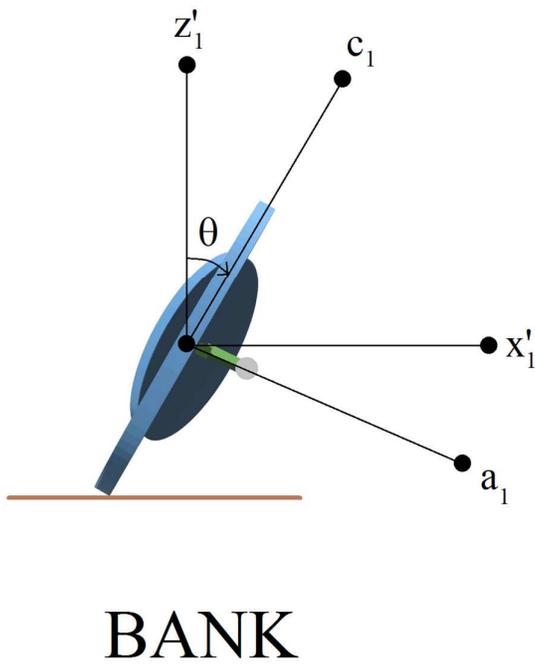

**BANK**

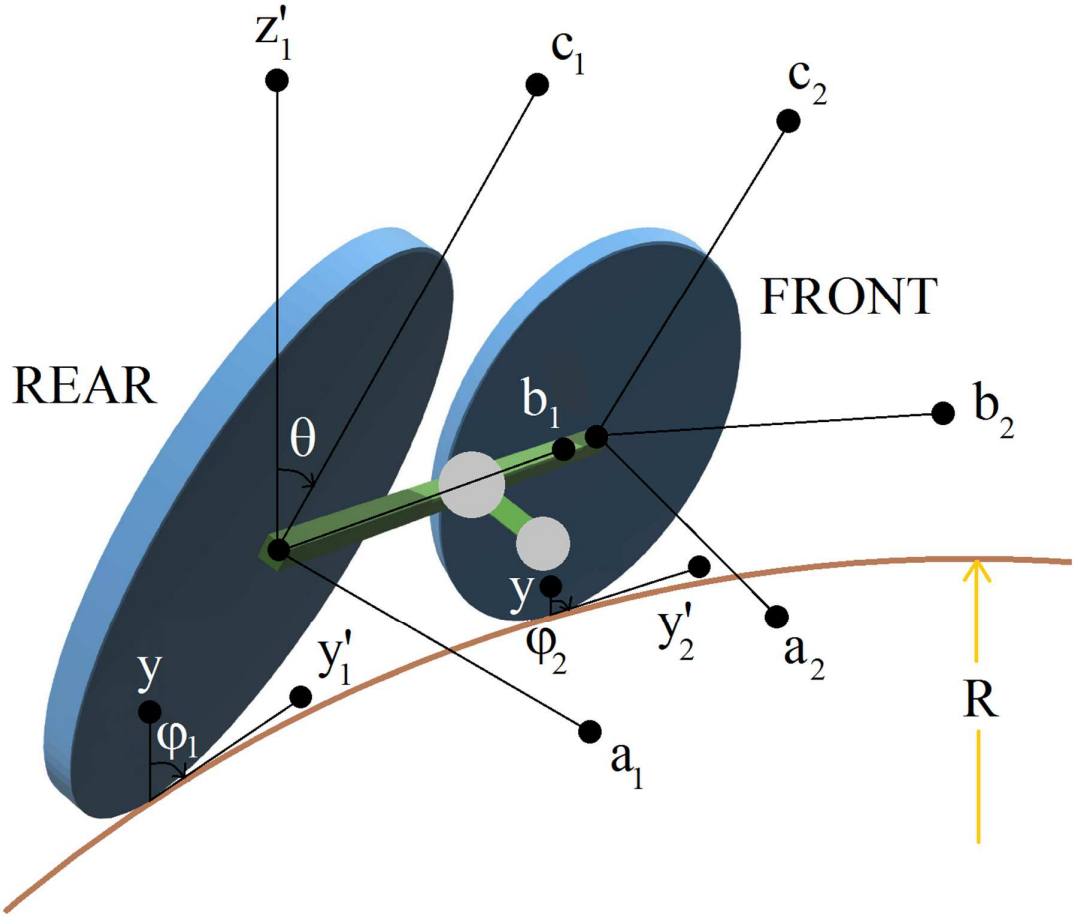

Figure 6 : *A three-dimensional view of the whole motorcycle. The brown curve on the ground shows the turning circle –* R *is its radius and its centre falls outside the diagram.*

Although this contradiction can be resolved by adding on an appropriate velocity to the truss CM, this velocity expression becomes quite complicated when the eccentric mass $m_4$ and the bank are taken into account. The easiest way around is to simply shift the origin of the truss to $O_1$ so that the translational motion arising from yaw angular velocity gets eliminated. When the truss comes up for torque balancing, we will figure out how to handle the effect of the non-inertial force on it with no great difficulty. (As an aside we note that this problem does not occur with the front wheel because that has the freedom to yaw relative to the truss.) Thus, for rotational purposes, both B1 and B3 will have their origins at $O_1$ and B2 will have its origin at $O_2$. Figures 5-6 show complete views of the mobike's geometry. At this point, the number of angle variables has come down from 9 to 5 – the surviving ones are $\varphi_1$, $\varphi_2$, $\theta$, $\psi_1$ and $\psi_2$.

The concept of rolling without slipping, which entered in the last but one paragraph, is now treated in greater detail. Under normal conditions (dry weather, new tyre, well-surfaced road) mobike wheels are designed to roll without slipping on the road. In our model, we will assume that rolling without slipping takes place always. This amounts to more constraints, this time on velocity rather than position. The most complete and rigorous definition of rolling without slipping of an object on a plane is that the component of the velocity of the contact point between the object and the plane parallel to the plane is identically zero. The additional (and very plausible) assumption that both the ground and the tyres are infinitely stiff and undergo no deformation on account of their mutual action guarantees that the velocity component of the contact point perpendicular to the ground plane is zero as well. Hence the rolling constraints are

$$\mathbf{v}_{C1/G} = 0 \quad , \tag{18a}$$

$$\mathbf{v}_{C2/G} = 0 \quad , \tag{18b}$$

where C denotes the contact points on both wheels, G denotes ground and $\mathbf{v}_{A/B}$ denotes the velocity vector of point A with respect to point B.

We now turn to expressing (18) in a form which will be of greater utility in the subsequent steps of the derivation. We have

$$\mathbf{v}_{C1/G} = \mathbf{v}_{O1/G} + \mathbf{v}_{C1/O1} \quad , \text{ or} \tag{19a}$$

$$0 = \mathbf{v}_{O1/G} + \boldsymbol{\omega}_1 \times \left(-r\hat{\mathbf{c}}_1\right) \quad . \tag{19b}$$

Now one component of $\mathbf{v}_{O1/G}$ arises from the translational speed $V$ of the mobike in the forward i.e. $b_1$ direction; a second component arises from the banking action. Thus,

$$V\hat{\mathbf{b}}_1 + \mathbf{v}_{\text{bank}} + r\left(\dot{\psi}_1 - \dot{\phi}_1\sin\theta\right)\hat{\mathbf{b}}_1 - r\dot{\theta}\hat{\mathbf{a}}_1 = 0 \quad , \tag{20}$$

wherefrom

$$\mathbf{v}_{\text{bank}} = r\dot{\theta}\hat{\mathbf{a}}_1 \quad , \tag{21a}$$

$$V = -r\left(\dot{\psi}_1 - \dot{\phi}_1\sin\theta\right) \quad . \tag{21b}$$

If the mobike is not accelerating then $V$ is a constant. Let us assume that this is the case here – the forward acceleration/de-acceleration of the mobike can be added on without much trouble and is not really an issue of fundamental interest. Then, (21b) gives a constraint between rear wheel spin and yaw rates while (21a) makes the (clear from geometry) statement that the banking action imparts to the CM a velocity in the $a_1$ direction. Note also the signs here – since $V$ is positive, $\dot{\psi}_1$ must be negative, and indeed from the geometry it is so.

Let us now take on the second rolling constraint, (18b). The first step is

$$\mathbf{v}_{C2/G} = \mathbf{v}_{O1/G} + \mathbf{v}_{O2/O1} + \mathbf{v}_{C2/O2} \quad . \tag{22}$$

We just evaluated the first term on this RHS; since the truss rotates about $O_1$, the second term can be written as $\boldsymbol{\omega}_3 \times \mathbf{r}_3$; the third term is of course $\boldsymbol{\omega}_2 \times \left(-r\hat{\mathbf{c}}_2\right)$. This simple form of the second term is an additional reason why $O_1$ is a good choice of origin for the truss; from this origin, the geometry shows that $\mathbf{r}_3 = 2l_3\hat{\mathbf{b}}_1$. Then, we can readily evaluate all the terms to get

$$r\dot{\theta}\hat{\mathbf{a}}_1 + V\hat{\mathbf{b}}_1 - 2l_3\dot{\phi}_1\cos\theta\,\hat{\mathbf{a}}_1 - 2l_3\dot{\phi}_1\sin\theta\,\hat{\mathbf{c}}_1 - r\dot{\theta}\hat{\mathbf{a}}_2 + r\left(\dot{\psi}_2 - \dot{\phi}_2\sin\theta\right)\hat{\mathbf{b}}_2 = 0 \quad . \tag{23}$$

The problem with this equation is that it is expressed in two different bases, so we now need the equation which performs the change of basis from $a_1,b_1,c_1$ to $a_2,b_2,c_2$. The transformation steps should be to first invert the bank, then apply the yaw $\Phi$ about $z$ and again apply the bank, giving the rotation matrix $\mathbf{Y}(\theta)\mathbf{Z}(\Phi)\mathbf{Y}(-\theta)$ or

$$\begin{bmatrix} \Gamma_{a2} \\ \Gamma_{b2} \\ \Gamma_{c2} \end{bmatrix} = \begin{bmatrix} \cos\theta & 0 & -\sin\theta \\ 0 & 1 & 0 \\ \sin\theta & 0 & \cos\theta \end{bmatrix} \begin{bmatrix} \cos\Phi & \sin\Phi & 0 \\ -\sin\Phi & \cos\Phi & 0 \\ 0 & 0 & 1 \end{bmatrix} \begin{bmatrix} \cos\theta & 0 & \sin\theta \\ 0 & 1 & 0 \\ -\sin\theta & 0 & \cos\theta \end{bmatrix} \begin{bmatrix} \Gamma_{a1} \\ \Gamma_{b1} \\ \Gamma_{c1} \end{bmatrix} \quad , \text{ or} \tag{24a}$$

$$\mathbf{R}_{a1b1c1}^{a2b2c2} = \begin{bmatrix} \cos\Phi\cos^2\theta + \sin^2\theta & \sin\Phi\cos\theta & \left(\cos\Phi - 1\right)\cos\theta\sin\theta \\ -\sin\Phi\cos\theta & \cos\Phi & -\sin\Phi\sin\theta \\ \left(\cos\Phi - 1\right)\cos\theta\sin\theta & \sin\Phi\sin\theta & \cos\Phi\sin^2\theta + \cos^2\theta \end{bmatrix} \quad . \tag{24b}$$

In (24a), $\mathbf{\Gamma}$ denotes an arbitrary vector. The inverse transformation is effected by its transpose i.e.

$$\mathbf{R}_{a2b2c2}^{a1b1c1} = \begin{bmatrix} \cos\Phi\cos^2\theta + \sin^2\theta & -\sin\Phi\cos\theta & \left(\cos\Phi - 1\right)\cos\theta\sin\theta \\ \sin\Phi\cos\theta & \cos\Phi & \sin\Phi\sin\theta \\ \left(\cos\Phi - 1\right)\cos\theta\sin\theta & -\sin\Phi\sin\theta & \cos\Phi\sin^2\theta + \cos^2\theta \end{bmatrix} \quad . \tag{25}$$

Using (25), I can write (23) entirely in the $a_1,b_1,c_1$ basis; this followed by some easy trigonometry gives the three componental equations

$$a_1 : r\dot{\theta}\left(1 - \cos\Phi\right)\cos\theta - 2l_3\dot{\phi}_1 - r\left(\dot{\psi}_2 - \dot{\phi}_2\sin\theta\right)\sin\Phi = 0 \quad , \tag{26a}$$

$$b_1 : V - r\dot{\theta}\sin\Phi\cos\theta + r\left(\dot{\psi}_2 - \dot{\phi}_2\sin\theta\right) = 0 \quad , \tag{26b}$$

$$c_1 : r\dot{\theta}\left(1 - \cos\Phi\right)\cos\theta - 2l_3\dot{\phi}_1 - r\left(\dot{\psi}_2 - \dot{\phi}_2\sin\theta\right)\sin\Phi = 0 \quad . \tag{26c}$$

The first equation above is identical to the third one hence only two of these are actual constraints.

These equations are exact (within the purview of our model) but are cumbersome and intractable for practical applications. Hence I will now invoke the fastness of the mobike. This assumption means that $\dot{\psi}_1, \dot{\psi}_2 \gg \dot{\phi}_1, \dot{\phi}_2, \dot{\theta}$. For a racing mobike this is certainly reasonable – during a turn, the yaw rate is probably one revolution in a minute, the maximum bank rate one revolution per second or lower, and the wheel spin rates ten to fifty times that. Under the aegis of the fastness assumption, we can rework the rolling constraints (21b) and (26b-c) as

$$r\dot{\psi}_1 = -V \quad , \tag{27a}$$

$$V - r\dot{\theta}\sin\Phi\cos\theta + r\dot{\psi}_2\cos\Phi = 0 \quad . \tag{27b}$$

$$r\dot{\theta}\left(1 - \cos\Phi\right)\cos\theta - 2l_3\dot{\phi}_1 - r\dot{\psi}_2\sin\Phi\cos\theta = 0 \quad . \tag{27c}$$

The first one of these implies that the rear wheel spin rate is a constant; since it is negative I will call it $-v_1$, where $v_1=V/r$. In the second and third equations if we substitute $\dot\theta=0$ i.e. the condition of a steady state turn, then we get

$$\dot\psi_2 = -\frac{v_1}{\cos\Phi} \quad , \tag{28a}$$

$$\dot\varphi_1 = \frac{rv_1\tan\Phi}{2l_3} \Rightarrow \tan\Phi = -\frac{2l_3\dot\varphi_1}{rv_1} \quad . \tag{28b}$$

But now, from fastness of the mobike, the term $\dot\varphi_1/v_1$ is small, implying that $\tan\Phi$ and hence $\Phi$ itself are also small. Thus, $\Phi$ in fact is a small variable whose linear dynamics should be sufficient. Although (28) is derived for a steady state turn, an angle which is 'small' in steady state cannot really become 'large' even in the dynamic condition.

It is noteworthy that the linearity in $\Phi$ arose naturally as a by-product of the constraints which the system has to satisfy and was not imposed by hand for the purpose of simplification. To get a better idea of this, let us use the fact that the radius of curvature of the turn (as we saw in the coin calculation) is $R=V/\dot\varphi_1$, which evaluates to

$$R = -\frac{2l_3}{\tan\Phi} \Rightarrow \tan\Phi = -\frac{2l_3}{R} \quad . \tag{29}$$

In a typical application, $2l_3$ is of the order of 1-2 m and $R$ is 30-100 m hence $\Phi$ is one tenth of a radian or less. For such a small angle, a linear approximation really works fine. In $\theta$, where nonlinearity is very much real (it can range from zero to 60º or more), I will keep the equations very much nonlinear.

With this in mind, I now substitute the small $\Phi$ assumption in (27). These yield relations for the rear wheel yaw rate and the front wheel spin rate in terms of $\Phi$; they can then be differentiated and the relation $\varphi_2=\varphi_1+\Phi$ used to get six ultimate constraint relations which I tabulate below :

$$\dot\varphi_1 = \frac{rv_1}{2l_3}\Phi \quad , \tag{30a}$$

$$\ddot\varphi_1 = \frac{rv_1}{2l_3}\dot\Phi \quad , \tag{30b}$$

$$\dot\varphi_2 = \dot\Phi + \frac{rv_1}{2l_3}\Phi \quad , \tag{30c}$$

$$\ddot\varphi_2 = \ddot\Phi + \frac{rv_1}{2l_3}\dot\Phi \quad , \tag{30d}$$

$$\dot\psi_2 = -v_1 + \Phi\dot\theta\cos\theta \quad , \tag{30e}$$

$$\ddot\psi_2 = \dot\Phi\dot\theta\cos\theta + \Phi\ddot\theta\cos\theta - \Phi\dot\theta^2\sin\theta \quad . \tag{30f}$$

Although the inclusion of the second term in $\dot{\psi}_2$ appears like a violation of the fastness approximation, it has to be kept because of its contribution to $\ddot{\psi}_2$, which will come up when we do the rotational equations.

Thus, the rolling without slipping constraint has enabled us to eliminate $\psi_1$, and express all of $\varphi_1$, $\varphi_2$ and $\psi_2$ in terms of $\Phi$. Out of the five variables which remained before invoking this constraint, only two are left now : $\theta$ and $\Phi$. These two thus become the variables in terms of which I will eventually write the equations of motion.

## 3 Acceleration and the equations of translation

To write Newton's laws we need expressions for acceleration of the CMs of B1, B2 and B3. The easiest is B1 because it is just like a coin moving in a circular trajectory. Copying (7b) without the gravity terms, we have

$$\alpha_{1a1} = \frac{V^2}{R_1}\cos\theta + r\ddot{\theta} \quad , \tag{31a}$$

$$\alpha_{1b1} = r\dot{\varphi}_1\dot{\theta}\cos\theta \quad , \tag{31b}$$

$$\alpha_{1c1} = \frac{V^2}{R_1}\sin\theta - r\dot{\theta}^2 \quad . \tag{31c}$$

Here I am using $\boldsymbol{\alpha}$ for linear acceleration because $a$ is an axis name. The subscript '$1a1$' means acceleration of B1, $a_1$ component and so on. These equations do not make use of the constraints (30) but I will save that substitution for the end – they look much more transparent this way.

The acceleration of $O_2$ is easy. Although (22) might imply a daisy chain of acceleration terms, we recognize that the front wheel, like the rear one, is just yawing, banking and pitching while its lowest point remains stationary. For the centripetal term, the appropriate radius of curvature will be $R_2 = V / \dot{\varphi}_2$, and the acceleration due to banking will have identical forms as (31) in the $a_2, b_2, c_2$ basis. Thus,

$$\alpha_{2a2} = \frac{V^2}{R_2}\cos\theta + m_2 r\ddot{\theta} \quad , \tag{32a}$$

$$\alpha_{2b2} = r\dot{\varphi}_2\dot{\theta}\cos\theta \quad , \tag{32b}$$

$$\alpha_{2c2} = \frac{V^2}{R_2}\sin\theta - r\dot{\theta}^2 \quad . \tag{32c}$$

What is comparatively non-trivial is the acceleration of the CM $O_3$ of the truss B3. Here, the direction of centripetal acceleration is not apparent a priori and I will resort to a first principles computation of the acceleration as a whole. The position vector of $O_3$ relative to $O_1$ is

$$\mathbf{r}_{O3/O1} = l_3\hat{\mathbf{b}}_1 + \frac{m_4}{m_3}l_4\hat{\mathbf{a}}_1 \qquad ,$$

$$= \eta l_4\hat{\mathbf{a}}_1 + l_3\hat{\mathbf{b}}_1 \tag{33}$$

where the second line acts as the definition of $\eta$. Taking the derivative,

$$\mathbf{v}_{O3/G} = \mathbf{v}_{O1/G} + \mathbf{v}_{O3/O1}$$

$$= r\dot{\theta}\hat{\mathbf{a}}_1 + V\hat{\mathbf{b}}_1 + \boldsymbol{\omega}_3 \times \mathbf{r}_{O3/O1} \qquad . \tag{34}$$

$$\left(r\dot{\theta} - l_3\omega_{3c1}\right)\hat{\mathbf{a}}_1 + \left(V + \eta l_4\omega_{3c1}\right)\hat{\mathbf{b}}_1 + \left(l_3\omega_{3a1} - \eta l_4\omega_{3b1}\right)\hat{\mathbf{c}}_1$$

For acceleration I need to differentiate this again; this time, since the above $\mathbf{v}$ is valid in the rotating dual of $a_1,b_1,c_1$ frame, the derivative must be material derivative i.e. $\mathrm{d}/\mathrm{d}t = \partial/\partial t + \boldsymbol{\omega}\times$. These two steps followed by trigonometric simplifications yield

$$\alpha_{3a1} = r\ddot{\theta} - l_3\ddot{\varphi}_1\cos\theta - \eta l_4\dot{\theta}^2 - V\dot{\varphi}_1\cos\theta - \eta l_4\dot{\varphi}_1^2\cos^2\theta \qquad , \tag{35a}$$

$$\alpha_{3b1} = \eta l_4\ddot{\varphi}_1\cos\theta - 2\eta l_4\dot{\varphi}_1\dot{\theta}\sin\theta - l_3\ddot{\varphi}_1^2 + r\dot{\varphi}_1\dot{\theta}\cos\theta \qquad , \tag{35b}$$

$$\alpha_{3a3} = -l_3\ddot{\varphi}_1\sin\theta - \eta l_4\ddot{\theta}_1 - V\dot{\varphi}_1\sin\theta - \eta l_4\dot{\varphi}_1^2\cos\theta\sin\theta - r\dot{\theta}^2 \qquad . \tag{35c}$$

With the accelerations on the table, the expressions for forces can now be written down.

The force exerted **on** body A **by** body B will be denoted as $\mathbf{F}_{AB}$. Further, since components of such forces will involve four subscripts, I will move the AB up to the superscript position. Hence $\mathbf{F}^{13}$ will mean the force exerted on the rear wheel by the truss, and not the thirteenth power of F. G of course means ground. The forces exerted on B1 are by the truss and the ground, the forces exerted on B2 are again by the truss and the ground, and the forces on B3 are by the two wheels. The forces from the ground consist of normal reaction as well as friction – since these together span the entire three-dimensional space I will not try to characterize them separately but lump them all into single vectors $\mathbf{F}^{1G}$ and $\mathbf{F}^{2G}$. The resultant force on each body must equal its mass times acceleration plus the reverse of gravity. This latter is $+mg\hat{\mathbf{z}}$ and can be split into $a,b,c$ components without trouble. This gives, for the rear wheel,

$$F_{a1}^{1G} + F_{a1}^{13} = m_1\alpha_{1a1} - m_1 g\sin\theta \qquad , \tag{36a}$$

$$F_{b1}^{1G} + F_{b1}^{13} = m_1\alpha_{1b1} \qquad , \tag{36b}$$

$$F_{c1}^{1G} + F_{c1}^{13} = m_1\alpha_{1c1} + m_1 g\cos\theta \qquad , \tag{36c}$$

where a componental notation is preferred to a vector one keeping in mind the future development. The front wheel has

$$F_{a2}^{2G} + F_{a2}^{23} = m_2\alpha_{2a2} - m_2 g\sin\theta \qquad , \tag{37a}$$

$$F_{b2}^{2G} + F_{b2}^{23} = m_2\alpha_{2b2} \qquad , \tag{37b}$$

$$F_{c2}^{2G} + F_{c2}^{23} = m_2\alpha_{2c2} + m_2 g\cos\theta \qquad . \tag{37c}$$

The truss satisfies

$$F_{a1}^{31} + F_{a1}^{32} = m_3\alpha_{3a1} - m_3 g\sin\theta \qquad , \tag{38a}$$

$$F_{b1}^{31} + F_{b1}^{32} = m_3 \alpha_{3b1} \quad , \tag{38b}$$

$$F_{c1}^{31} + F_{c1}^{32} = m_3 \alpha_{3c1} + m_3 g \cos\theta \quad . \tag{38c}$$

Newton's Third Law of Motion says

$$\mathbf{F}^{13} = -\mathbf{F}^{31} \quad , \tag{39a}$$

$$\mathbf{F}^{23} = -\mathbf{F}^{32} \quad . \tag{39b}$$

This completes the content of this Section.

Although no explicit expression for any force has been obtained, further progress cannot be made at this point. It is the torque balance which will provide concrete expressions for the various forces. The same phenomenon also occurs in any 2D system with reaction or friction, like say a lever supported at two points. Even though the system here is vastly more complicated, the basic principle still remains the same.

## 4  Angular velocity and the equations of rotation

Here I will write down the rotational equations for each of the bodies B1, B2 and B3. Combining those with (36-8) should somehow lead to an overall EOM of the mobike. For each wheel I will use the symmetric Euler equation (6). Substituting the angular velocity terms leads to

$$I_a \left( -\ddot{\varphi}\sin\theta - \dot{\varphi}\dot{\theta}\cos\theta + \ddot{\psi} \right) = T_a \quad , \tag{40a}$$

$$I_s \ddot{\theta} + I_a \left( \dot{\psi} - \dot{\varphi}\sin\theta \right)\dot{\varphi}\cos\theta + I_s \dot{\varphi}^2 \cos\theta\sin\theta = T_b \quad , \tag{40b}$$

$$I_s \left( \ddot{\varphi}\cos\theta - 2\dot{\varphi}\dot{\theta}\sin\theta \right) - I_a \dot{\theta}\left( \dot{\psi} - \dot{\varphi}\sin\theta \right) = T_c \quad , \tag{40c}$$

where $I_a$ denotes the axial moment of inertia and $I_s$ the other (in-plane) one. Under the fastness assumption, many terms drop out; further for B1, $\dot{\psi}_1$ is a constant and its derivative is zero. Thus, the LHSes of torque balance for B1 and B2 are within reach. For B3, $a_1,b_1,c_1$ is actually the full body frame and its rotational equation is Euler's equation as commonly known; I will perform the substitution (3) in a while. But what guarantees that $a_1,b_1,c_1$ is principal ?

In my model, the guarantee comes from the simple distribution of the masses on the truss. The point mass model of frame and driver did not enter the picture while writing (35) – a more complicated mass distribution would have produced a similar-looking result. But it might also have changed the principal basis for the truss about $O_1$ from $a_1,b_1,c_1$ to some intractably oriented coordinate set, and that would have now resulted in an enormous amount of extra effort with no commensurate reward. In a real mobike, the design symmetries would ensure that $a_1,b_1,c_1$ is more or less principal, and one can always add on the refinements later.

Now that we are more or less confident about the LHSes, let us turn to the RHSes. The forces acting on B1 are $\mathbf{F}^{1G}$ at $C_1$, and $\mathbf{F}^{13}$ and gravity at $O_1$. The torque of the latter two about $O_1$ will be trivially zero and the first one should also be tractable. But, the truss and the wheel can also exert equal and opposite torques on each other with no force, just as a wall being drilled exerts an enormous resistive torque on the drill bit but does not cause it to (tend to) move this way and that. Can there be such an unknown torque along every body's every axis in addition to all the existing unknown forces ?

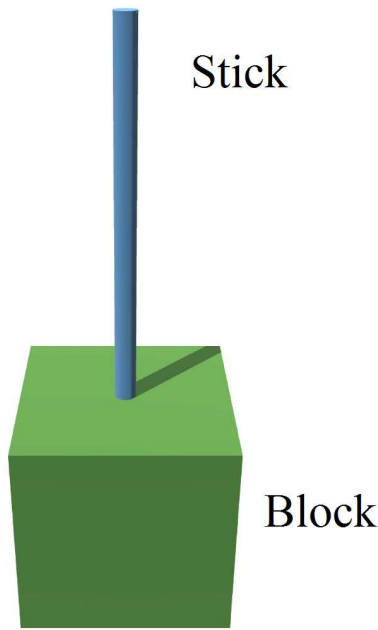

Figure 7 : *A stick attached to a heavy block through a bearing. We catch hold of the stick and try to rotate the apparatus. Clearly, the block responds with a force-free torque only when the rotation is about the axis of the stick.*

To resolve this issue, we imagine the simplest example of a pivoted system, where a long stick is connected using a bearing to the centre of a heavy rigid block. This is shown in Fig. 7. We catch the stick at its centre and apply a torque on it. Now, from this Figure it is clear that if we apply a torque along the axis of the stick, the block will respond with a counter-torque and resist being rotated. But the centre of neither the stick nor the block will tend to move, implying that this torque has been generated without a resultant force. However if we try applying a torque perpendicular to the axis of the stick, then the block will not only resist us but also move along with the stick. Clearly, in this case, the block and the stick interact with a mutual force, which also has some non-zero moment. If we have already accounted for the force, then we do not need to consider the moment again – that will be an extra floating variable which will never get determined. The conclusion of this simple example is that a bearing can exert a force-free torque only along an axis about which a rotation does not (or hypothetical rotation would not) lead of motion of the CMs of any of the constituent elements. In the mobike situation, the wheel-truss bearings are more complicated, but the principle remains the same. Considering the rear wheel bearing $O_1$, the rotation of the wheel about $a_1$-axis does not cause any CM displacement while a hypothetical rotation of the truss about $b_1$ would also not cause any CM displacement if $\eta=0$ (this rotation is forbidden but that's another story). Thus this bearing can exert force-free torques about $a_1$ and $b_1$-axes. As for the front wheel, the bearing at $O_2$ features no CM displacements for rotation of the wheel about $a_2$-axis and hypothetical rotation of the truss about $b_1$-axis, hence its force-free torque axes must be $a_2$ and $b_1$. On the other hand, the steering of the front wheel leads to yaw of the whole mobike and displacement of both truss and wheel CMs so a torque in this direction does come with a force attached. If these assumptions on the bearings work then fine, and if they lead to a contradiction, we can always go back and correct them.

So now let us write the rotational equation for B1. RHS first, the torque due to $\mathbf{F}^{1G}$ is

$$
\begin{aligned}
\mathbf{r} \times \mathbf{F}^{1G} &= \left(-r\hat{\mathbf{c}}_1\right) \times \mathbf{F}^{1G} \\
&= rF_{b1}^{1G}\hat{\mathbf{a}}_1 - rF_{a1}^{1G}\hat{\mathbf{b}}_1
\end{aligned} \tag{41}
$$

Letting the force-free bearing torque be

$$
\mathbf{T}^{13} = \lambda\hat{\mathbf{a}}_1 + \mu\hat{\mathbf{b}}_1 \quad , \tag{42}
$$

I now take (40), apply fastness and constancy of $v_1$ and write

$$
I_{1a}\left(-\ddot{\phi}_1\sin\theta - \dot{\phi}_1\dot{\theta}\cos\theta\right) = rF_{b1}^{1G} + \lambda \quad , \tag{43a}
$$

$$
I_{1s}\ddot{\theta} - I_{1a}v_1\dot{\phi}_1\cos\theta = -rF_{a1}^{1G} + \mu \quad , \tag{43b}
$$

$$
I_{1s}\ddot{\phi}_1\cos\theta + I_{1a}v_1\dot{\theta} = 0 \quad . \tag{43c}
$$

Imposing the constraints (30) leads to the final form of the equations for B1 :

$$
\text{"B1/1"} : I_{1a}\left(-\frac{rv_1}{2l_3}\dot{\Phi}\sin\theta - \frac{rv_1}{2l_3}\Phi\dot{\theta}\cos\theta\right) = rF_{b1}^{1G} + \lambda \quad , \tag{44a}
$$

$$
\text{"B1/2"} : I_{1s}\ddot{\theta} - I_{1a}\frac{rv_1^2}{2l_3}\Phi\cos\theta = -rF_{a1}^{1G} + \mu \quad , \tag{44b}
$$

$$
\text{"B1/3"} : I_{1s}\frac{rv_1}{2l_3}\dot{\Phi}\cos\theta + I_{1a}v_1\dot{\theta} = 0 \quad . \tag{44c}
$$

Equation (44c) is in fact one of the two components of the ultimate EOM of the motorcycle.

This statement might appear incredible at first because a lot of work is still obviously left – the front wheel and the truss are yet to be analysed. But it is true because we have found one equation in $\theta$ and $\Phi$, (44c), which has zero on the RHS. Since our system has two variables, two fully determinate (differential) equations connecting them will be sufficient for our purposes. Equation (44c) is one of them; it clearly says that some function of $\theta$ and $\Phi$ (and their derivatives) equals zero, and not some unknown force or torque. So by default, this is one of the two ultimate equations of motion of the motorcycle. It is nonlinear, but quite simple in structure.

Now for the torque equation of the front wheel B2. As with B1, gravity and $\mathbf{F}^{23}$ have no torques about $O_2$, while $\mathbf{F}^{2G}$ has $rF_{b2}^{2G}\hat{\mathbf{a}}_2 - rF_{a2}^{2G}\hat{\mathbf{b}}_2$. By our assumption on the bearing, the force-free torque between front wheel and truss has to be

$$
\mathbf{T}^{23} = \sigma\hat{\mathbf{a}}_2 + \tau\hat{\mathbf{b}}_1 \quad . \tag{45}
$$

But this is in a heterogeneous basis. To get it in any one basis, we need to use (24) or (25) on the non-conforming component, after modifying them to account for the smallness of $\Phi$. Applying fastness on (40) and expressing (45) in the 2-basis,

$$I_{2a}\left(-\ddot{\varphi}_2\sin\theta-\dot{\varphi}_2\dot{\theta}\cos\theta+\ddot{\psi}_2\right)=rF_{b2}^{2G}+\sigma+\tau\Phi\cos\theta \quad, \tag{46a}$$

$$I_{2s}\ddot{\theta}+I_{2a}\dot{\psi}_2\dot{\varphi}_2\cos\theta=-rF_{a2}^{2G}+\tau \quad, \tag{46b}$$

$$I_{2s}\ddot{\varphi}_2\cos\theta-I_{2a}\dot{\theta}\dot{\psi}_2=\tau\Phi\sin\theta \quad. \tag{46c}$$

Then applying (30),

$$\text{"B2/1" :}\ I_{2a}\left[\begin{array}{l}\left(-\ddot{\Phi}-\dfrac{r\nu_1}{2l_3}\Phi\right)\sin\theta-\left(\dot{\Phi}+\dfrac{r\nu_1}{2l_3}\Phi\right)\dot{\theta}\cos\theta\\[2mm]+\dot{\Phi}\dot{\theta}\cos\theta+\Phi\ddot{\theta}\cos\theta-\Phi\dot{\theta}^2\sin\theta\end{array}\right]=rF_{b2}^{2G}+\sigma+\tau\Phi\cos\theta \quad, \tag{47a}$$

$$\text{"B2/2" :}\ I_{2s}\ddot{\theta}+I_{2a2}\left(-\nu_1+\Phi\dot{\theta}\cos\theta\right)\left(\dot{\Phi}+\dfrac{r\nu_1}{2l_3}\Phi\right)\cos\theta=-rF_{a2}^{2G}+\tau \quad, \tag{47b}$$

$$\text{"B2/3" :}\ I_{2s}\left(\ddot{\Phi}+\dfrac{r\nu_1}{2l_3}\Phi\right)\cos\theta-I_{2a}\dot{\theta}\left(-\nu_1+\Phi\dot{\theta}\cos\theta\right)=\tau\Phi\sin\theta \quad. \tag{47c}$$

Although it might appear that I have made inadequate use of fastness in the above, especially in (47c), that is not the case as we will see in a while.

The last set of torque equations is for the truss B3. Here the origin $O_1$ is not the CM of the truss and neither is it a stationary point [we calculated its acceleration in (31)]. So now we go into a frame which shares the acceleration $\boldsymbol{\alpha}_1$ of $O_1$, add on the appropriate non-inertial force to $O_3$ and then calculate the torque of that force along with all the other forces. Thus the forces on the truss are : $\mathbf{F}^{31}$ at $O_1$, whose torque is zero, gravity $-m_3 g\hat{\mathbf{z}}$ and the non-inertial force $-m_3\boldsymbol{\alpha}_1$ at $O_3$ and $\mathbf{F}^{32}$ at $O_2$, the latter three all contributing to the torque. Then there are the force-free torques $\mathbf{T}^{31}=-\lambda\hat{\mathbf{a}}_1-\mu\hat{\mathbf{b}}_1$ from the rear wheel and $\mathbf{T}^{32}=-\sigma\hat{\mathbf{a}}_2-\tau\hat{\mathbf{b}}_1$ from the front wheel. The torques of the forces on $O_3$ can be calculated as

$$\begin{aligned}\mathbf{T}&=\mathbf{r}\times\mathbf{F}\\&=\left(\eta l_4\hat{\mathbf{a}}_1+l_3\hat{\mathbf{b}}_1\right)\times\left[\left(m_3 g\sin\theta-m_3\alpha_{1a1}\right)\hat{\mathbf{a}}_1-m_3\alpha_{1b1}\hat{\mathbf{b}}_1-\left(mg\cos\theta+m\alpha_{1c1}\right)\hat{\mathbf{c}}_1\right]\\&=-m_3 l_3\left(g\cos\theta+\alpha_{1c1}\right)\hat{\mathbf{a}}_1+\eta m_3 l_4\left(g\cos\theta+\alpha_{1c1}\right)\hat{\mathbf{b}}_1+\\&\quad\left[-\eta m_3 l_4\alpha_{1b1}-m_3 l_3\left(g\sin\theta-\alpha_{1a1}\right)\right]\hat{\mathbf{c}}_1\end{aligned} \tag{48}$$

The torque of $\mathbf{F}^{32}$ is

$$\begin{aligned}\mathbf{T}&=\left(2l_3\hat{\mathbf{b}}_1\right)\times\mathbf{F}^{32}\\&=2l_3 F_{c1}^{32}\hat{\mathbf{a}}_1-2l_3 F_{a1}^{32}\hat{\mathbf{c}}_1\end{aligned} \tag{49}$$

The force-free torques can be projected into the $a_1,b_1,c_1$ basis using (25). Putting all these together and writing the LHS as the canonical Euler's equation with the substitution (3),

$$I_{3a1}\left(-\ddot{\varphi}_1\sin\theta-\dot{\varphi}_1\dot{\theta}_1\cos\theta\right)+\left(I_{3c1}-I_{3b1}\right)\dot{\varphi}_1\dot{\theta}\cos\theta=-m_3 l_3\left(g\cos\theta+\alpha_{1c1}\right)\\+2l_3 F_{c1}^{32}-\lambda-\sigma \quad, \tag{50a}$$

$$I_{3b1}\ddot{\theta}+\left(I_{3a1}-I_{3c1}\right)\left(-\dot{\varphi}_1^2\cos\theta\sin\theta\right)=\eta m_3 l_4\left(g\cos\theta+\alpha_{1c1}\right)-\mu-\sigma\Phi\cos\theta-\tau \quad, \tag{50b}$$

$$I_{3c1}\left(\ddot{\varphi}_1\cos\theta - \dot{\varphi}_1\dot{\theta}_1\sin\theta\right) + \left(I_{3b1} - I_{3a1}\right)\left(-\dot{\varphi}_1\dot{\theta}\sin\theta\right) = -\eta m_3 l_4 \alpha_{1b1}$$
$$-m_3 l_3 \left(g\sin\theta - \alpha_{1a1}\right) - 2l_3 F_{a1}^{32}. \tag{50c}$$

Finally, substituting the constraints (30),

$$\text{"B3/1"}: I_{3a1}\left(-\frac{rv_1}{2l_3}\dot{\Phi}\sin\theta\right) + \left(I_{3c1} - I_{3b1} - I_{3a1}\right)\frac{rv_1}{2l_3}\Phi\dot{\theta}\cos\theta,$$
$$= m_3 l_3\left(g\cos\theta + \alpha_{1c1}\right) + 2l_3 F_{c1}^{32} - \lambda - \sigma \tag{51a}$$

$$\text{"B3/2"}: I_{3b1}\ddot{\theta} + \left(I_{3a1} - I_{3c1}\right)\left(-\frac{r^2 v_1^2}{4l_3^2}\Phi^2\cos\theta\sin\theta\right),$$
$$= \eta m_3 l_4\left(g\cos\theta + \alpha_{1c1}\right) - \mu - \sigma\Phi\cos\theta - \tau \tag{51b}$$

$$\text{"B3/3"}: I_{3c1}\frac{rv_1}{2l_3}\dot{\Phi}\cos\theta + \left(I_{3a1} - I_{3b1} - I_{3c1}\right)\frac{rv_1}{2l_3}\Phi\dot{\theta}\sin\theta.$$
$$= \eta m_3 l_4 \alpha_{1b1} - m_3 l_3\left(g\sin\theta - \alpha_{1a1}\right) - 2l_3 F_{a1}^{32} \tag{51c}$$

These are the final form of the rotational equations of the truss.

The nine force balance equations (36-8) and the eight torque balance equations B1/1-B3/3 excluding the standalone B1/3 constitute an implicit equation of motion of the motorcycle – implicit because of their dependence on unknown forces and torques. In the next Section we will see how to convert them into an explicit EOM.

# 5  The Equations of motion and analysis of solutions

The first thing to note is that not all of the 17 force and torque equations may be required to obtain the second EOM. Some of those 17 might contain an EOM between them, while others can be prescriptions which determine the various components of force and torque in terms of the now known variables $\theta$ and $\Phi$. Inspection of the system is the only way forward from this point. What is reassuring however is that there is a match between the number of equations and the number of unknowns. The equation tally is of course 17; the unknown force and torque components are 3 each for $\mathbf{F}^{1G}$, $\mathbf{F}^{13}$, $\mathbf{F}^{23}$, $\mathbf{F}^{2G}$, and $\lambda$, $\mu$, $\sigma$, $\tau$ for a total of 16. Thus, 16 of the 17 equations will go into finding their values and the remaining one will be the second desired EOM.

We start from the one known EOM, B1/3. B2/3 clearly has some overlap with this one; cancelling off the common terms will yield $\tau$. (Here and henceforth, 'yield' will mean that the unknown force or torque component will get expressed in terms of the basic variables and their derivatives.) Substituting $\tau$ into B2/2 will give us $F_{a2}^{2G}$. A chain of force balances can then yield expressions for the $a_2$ components of $\mathbf{F}^{23}$, $\mathbf{F}^{13}$ and $\mathbf{F}^{1G}$ but these are of uncertain utility as they do not feature directly in the equations. In a parallel development, B3/3 can yield $F_{a1}^{32}$; successive force balances then produce expressions for the $a_1$ components of $\mathbf{F}^{2G}$, $\mathbf{F}^{13}$ and $\mathbf{F}^{1G}$.

Now however every remaining equation seems to feature not one but two unknown force and torque components on the RHS.

The lifting of this block is the last source of subtlety in the entire analysis. The eliminations so far have given us the $a_1$ and $a_2$ components of $\mathbf{F}^{2G}$, while B2/1 requires its $b_2$ component. The partial information of the $a_1$ and $a_2$ components (which are non-orthogonal) will give us $b_2$ if and only if the three unit vectors $\hat{\mathbf{a}}_1$, $\hat{\mathbf{a}}_2$ and $\hat{\mathbf{b}}_2$ are coplanar. To determine coplanarity we do the triple product test :

$$
\begin{aligned}
\hat{\mathbf{a}}_1 \cdot \left( \hat{\mathbf{a}}_2 \times \hat{\mathbf{b}}_2 \right) &= \hat{\mathbf{a}}_1 \cdot \hat{\mathbf{c}}_2 \\
&= \hat{\mathbf{a}}_1 \cdot \left( \Phi \sin\theta \hat{\mathbf{b}}_1 + \hat{\mathbf{c}}_1 \right) \quad , \\
&= 0
\end{aligned}
\tag{52}
$$

proving that the three vectors are indeed coplanar. Thus, $F_{b2}^{2G}$ is a known quantity. Substituting this and $\tau$ into B2/1 will yield $\sigma$, that into B3/2 give $\mu$ and $\mu$ and $F_{a1}^{1G}$ into B1/2 will result in the second EOM.

The execution of the above steps is just tedious but straightforward algebra, which need not be shown. The fastness approximation can be used liberally but must be invoked carefully. The safest algorithm is to classify all terms according to the size of their constituent angular velocities, and then keep the largest order. Individually, the large term is the spin rate $\nu_1$ while the bank rate $\dot\theta$ and the yaw rates $\dot\varphi_1, \dot\varphi_2$ and $\dot\Phi$ are all of comparable size and one order of magnitude smaller than $\nu_1$. Terms with second derivatives survive by default, as they should – we have no idea about the sizes of the 'angular accelerations'. We note also that $\nu_1 \Phi \simeq \dot\varphi$ hence $\Phi$ effectively 'neutralizes' $\nu_1$. Thus in an expression like $\nu_1 \dot\Phi \cos\theta + \nu_1 \dot\Phi \dot\theta \cos\theta \sin\theta + \ddot\theta$, the first term is of size spin times yaw, the second is yaw times bank while the third is an acceleration. Clearly, the second term is an order of magnitude smaller than the first, and can be dropped.

Doing the algebra now leads to the EOM of the motorcycle :

$$
\begin{aligned}
&\left[ I_{1s} + I_{2s} + I_{3b1} + (m_1 + m_2 + m_3) r^2 \right] \ddot\theta - \left[ \frac{(I_{1a} + I_{2a}) r + (m_1 + m_3) r^3}{2 l_3} \right] \nu_1^2 \Phi \cos\theta \\
&+ \frac{\eta m_3 l_4 r^2}{2 l_3} \nu_1^2 \Phi \sin\theta - \frac{m_2 r^3}{2 l_3} \nu_1^2 \Phi - (m_1 + m_2 + m_3) rg \sin\theta - \eta m_3 l_4 g \cos\theta \\
&- \frac{m_3 r^2}{2} \nu_1 \dot\Phi \cos\theta - m_2 r^2 \nu_1 \dot\Phi = 0
\end{aligned}
\tag{53}
$$

&

$$
\frac{I_{1s} r}{2 l_3} \dot\Phi \cos\theta + I_{1a} \dot\theta = 0 \quad .
\tag{54}
$$

This completes the derivation of the EOM.

A first observation is comparison with the coin EOM in the fast limit, (11). The structures of these EOMs are in fact strikingly similar with $\Omega$ of the coin corresponding to $\Phi$ of the mobike. For the coin, the first equation says that $\ddot{\theta}$ has a large negative $\cos\theta$ term balanced by a large positive $\sin\theta$ term. Here there is a similar structure (recall that $\Phi$ is generally negative). There are some additional terms arising from the complex geometry, but not all that many. The second equations for both systems are almost identical; paradoxically, (54) has an even simpler structure than (11b).

Fixed points of the system occur when $\dot{\theta} = \dot{\Phi} = 0$. The second equation vanishes while the first gives a relation between $\Phi$ and the equilibrium lean angle $\theta^*$. Note that for every $\Phi$ there is one $\theta^*$; the complete three-dimensional (in the sense of number of variables, not spatial dimensions) system (53-4) thus has a line of fixed points in the $\theta, \Phi$ plane. Since $\Phi$ is related by (29) to the radius of curvature, (53) predicts the steady state lean angle given the turn radius, the speed and the various parameters. Using that $\Phi = -2l_3/R$ and $V = v_1 r$, this angle $\theta^*$ satisfies the relation

$$\left[ \frac{1}{R} \left( \frac{I_{1a} + I_{2a}}{r} + (m_1 + m_3)r \right) \right] V^2 \cos\theta^* - \frac{\eta m_3 l_4}{R} V^2 \sin\theta^* +$$
$$\frac{m_2 r}{R} V^2 - (m_1 + m_2 + m_3) rg \sin\theta^* - \eta m_3 l_4 g \cos\theta^* = 0 \qquad (55)$$

Now for a realistic mobike it generally happens that the bulk of the mass comes from the frame, the engine and the driver i.e. $m_3 \gg m_1, m_2$. Typical values can be $m_1, m_2 = 5$ kg and $m_3 = 300$ kg. Hence, I can very reasonably keep only $m_3$ terms in (55) and drop all others. If I now assume that $\eta = 0$ i.e. the driver does not lean out of the mobike then that leads to the unbelievable answer

$$\theta^* = \arctan\frac{V^2}{gR} \qquad , \qquad (56)$$

same as (xx). Experiments also bring out the truth of (56); that is why it appears in the literature without being challenged.

Nevertheless, the 2D model leading to (xx) is still not credible as it predicts that the mobike is unstable (so do the Literature 3D models but we let that pass). Hence we now evaluate the stability of (56) as per (53-4), which we rewrite as

$$A\ddot{\theta} - B\Phi\cos\theta + C\Phi\sin\theta - D\Phi - E\sin\theta - F\cos\theta - G\dot{\Phi}\cos\theta - H\dot{\Phi} = 0 \qquad , \qquad (57a)$$
$$M\dot{\Phi}\cos\theta + P\dot{\theta} = 0 \qquad . \qquad (57b)$$

The definitions of $A$, $B$ etc. should be clear from the correspondence and we note that all of these are positive. Letting $\theta = \theta^* + X$, $\dot{\theta} = \dot{X} = Y$ and $\Phi = \Phi^* + z$, linear stability analysis yields

$$A\ddot{Y} + \dot{Y}\left[ \frac{P(G\cos\theta^* + H)}{M\cos\theta^*} \right] + Y\left[ \begin{array}{l} B\Phi^*\sin\theta^* + C\Phi^*\cos\theta^* - E\cos\theta^* + F\sin\theta^* + \\ \frac{P}{M\cos\theta^*}(B\cos\theta^* - C\sin\theta^* + D) \end{array} \right] \qquad (58)$$
$$= 0$$

First recall that $\Phi^*$ is negative if $\theta^*$ is positive. Then, the term on the first derivative obviously constitutes a positive damping; of the various terms in the coefficient of $Y$, $PB/M$ contributes the maximum positivity while $E\cos\theta^*$ contributes the maximum negativity. Since $B$ depends on speed, at sufficiently high speed (58) should describe a damped harmonic oscillator and not a repeller of some form. Numerical work confirms this for a typical mobike at a typical operating point. Hence the fixed point as per the original EOM (53-4) is stable.

One eigenvalue of the third order system is zero (the one for $X$); the other two describe damped harmonic oscillator. Because of this dynamics, perturbations in lean angle velocity and steering angle die out quickly, explaining the mobike's imperturbability on turns. The primary stiffness terms $B$ and $D$ get larger as speed increases. Hence the same perturbation, say from wind or from a deformity in the road surface, will cause a smaller and smaller disturbance as the mobike gets faster. The damping aids the stability even more. Note that the coefficients $G$ and $H$ also increase with speed. In fact, in many regions of parameter space and operating speed, (58) describes an overdamped oscillator. So during a high speed turn if the mobike suffers a perturbation, not only will it deviate a very small distance but it also won't keep oscillating about its fixed point. This is why the racing drivers can afford to take turns with apparently hair's breadth clearance.

We now analyse the source of discord between my above findings and the entire research literature. A physical answer to this may be found in SOMMERFELD who has adopted a Newtonian approach and also given some interpretation of his results. He argues that at high speed, the plane of the rear wheel tends to coincide with that of the front wheel, and the bicycle behaves as though the two wheels were rigidly bound. In the language of the present analysis, it means that $\Phi$ is zero. We already saw in (29) that $\Phi$ is small; we now quantify its smallness. The centripetal acceleration for the turn is provided by friction from the ground, which is bounded by the coefficient of friction. As a ballpark estimate let us take its value to be 1. Then, $V^2/R$ must equal $g$; a tighter turn will not be possible. Thus the turning radius increases quadratically as the velocity; since $\Phi = -2l_3/R$, its value goes as the inverse square of velocity. At $V$=130 km/hr, $R$ comes out as 130 m; if wheelbase $2l_3$=2 m then $\Phi$ is 1/60 of a radian or one degree.

It is really tempting to ignore $\Phi$ altogether – what difference can one degree make. Let us see what happens if we actually set $\Phi$ to zero in (53-4). This describes a straight mobike; almost all the terms in (53) and whole (54) disappear leaving behind what is clearly an unstable system. In the turning case, a derivation bypassing $\Phi$ will independently yield $V^2/R$ in place of $rv_1^2\Phi$ in (53); the terms involving $\dot{\Phi}$ will vanish. More importantly, (54) will also vanish. Now we saw that the primary source of stability in the linearized equation (58) came from (54) – in the absence of that equation, the dynamics would become unstable. This instability would of course come with a huge positive eigenvalue, and not the small one which the literature finds, but the example shows that $\Phi$ is by no means an ignorable variable at any speed. I suspect that an incorrect treatment of $\Phi$ is what has happened in SOMMERFELD's work. The later derivations are less transparent on account of their use of the Lagrangian, but since their instability is of identical nature, the phenomenon must be same.

A second reason behind this spurious instability is more mathematical. It lies in the restriction of all the works to ad hoc linearization rather than linearization of full nonlinear equations. In our analysis, one eigenvalue came out exactly zero, which in this context is very much plausible. The zero arises from the fact that there is a line of fixed points in the $\theta,\Phi$ plane – the eigenvalue for perturbation along that line must be zero. The physical interpretation of this line is that there is a lean angle for every turning radius and the mobike has no intrinsic preference for any one particular radius. This zero is what comes out as small positive in the literature. In the absence of a nonlinear equation, a zero eigenvalue is extremely hard to get correctly; the error in this instance happens to be on the positive side, resulting in a physically unjustifiable scenario.

As an indicator of the accuracy of this model, we note that the minimum speed at which a typical mobike becomes stable as per (58) is approximately 30 km/hr. This is in good agreement with what the various References have obtained. However, this model retains stability of the mobike at all higher speeds, as is observed in reality.

We will now use this model to quantitatively explain the phenomenon of counter-steering, which happens during entry into a turn. When a straight and speeding mobike approaches a curve, the driver is always instructed to initiate the turn by briefly steering opposite to the intended direction of turn and simultaneously generating a leaning torque with his hands and knees. The mobike initially turns the wrong way, and then leans in and starts turning the right way. To model this phenomenon, we need to modify our EOM and include provisions for torques exerted on the system by the driver (the EOM so far assumes that the driver is passive). There are two ways the additional torques can be incorporated – by a rigorous insertion in the 18 translation and rotational equations and performing the elimination afresh, or by using a physically informed 'trick'. Here I will go with the latter approach.

The trick is to recognize that the final EOM is essentially one equation of torque balance about $b$-axis (53) and about $c$-axis (54). Since $\Phi$ is small, we do not really care whether it is $b_1,c_1$ or $b_2,c_2$. Then, a $b$-axis torque applied by the driver will add on a term $T_{Db}$ to the RHS of (53) while a $c$-axis torque by driver will on a term $T_{Dc}$ to the RHS of (54). The modified equations in the presence of the driver are thus

$$
\left[ I_{1s} + I_{2s} + I_{3b1} + \left( m_1 + m_2 + m_3 \right) r^2 \right] \ddot{\theta} - \left[ \frac{\left( I_{1a} + I_{2a} \right) r + \left( m_1 + m_3 \right) r^3}{2l_3} \right] v_1^2 \Phi \cos \theta
$$
$$
+ \frac{\eta m_3 l_4 r^2}{2l_3} v_1^2 \Phi \sin \theta - \frac{m_2 r^3}{2l_3} v_1^2 \Phi - \left( m_1 + m_2 + m_3 \right) rg \sin \theta - \eta m_3 l_4 g \cos \theta \qquad , \qquad (59)
$$
$$
- \frac{m_3 r^2}{2} v_1 \dot{\Phi} \cos \theta - m_2 r^2 v_1 \dot{\Phi} = T_{Db}
$$

&

$$
\frac{I_{1s} r}{2l_3} \dot{\Phi} \cos \theta + I_{1a} \dot{\theta} = T_{Dc} \qquad . \qquad (60)
$$

To take a turn, the driver wants to get into the correct lean angle; the strongly stable fixed point ensures that Φ follows the lean angle closely. Qualitatively (see the rear view in Fig. 5), the angular momentum of the wheels is along the −*a*-axis; if the mobike starts leaning inwards for a starboard turn, then the change in **L** becomes along the +*c*-axis. Hence a positive *c*-axis torque by driver is required to get a positive lean angular velocity. Accordingly, we simulate (59-60) with the initial condition θ=Φ=0 and $\dot{\theta} = 0$ i.e. the bike is going straight. At *t*=0, the driver applies a constant positive $T_{Dc}$ and holds it for 1 time unit before releasing the torque. The results are shown below.

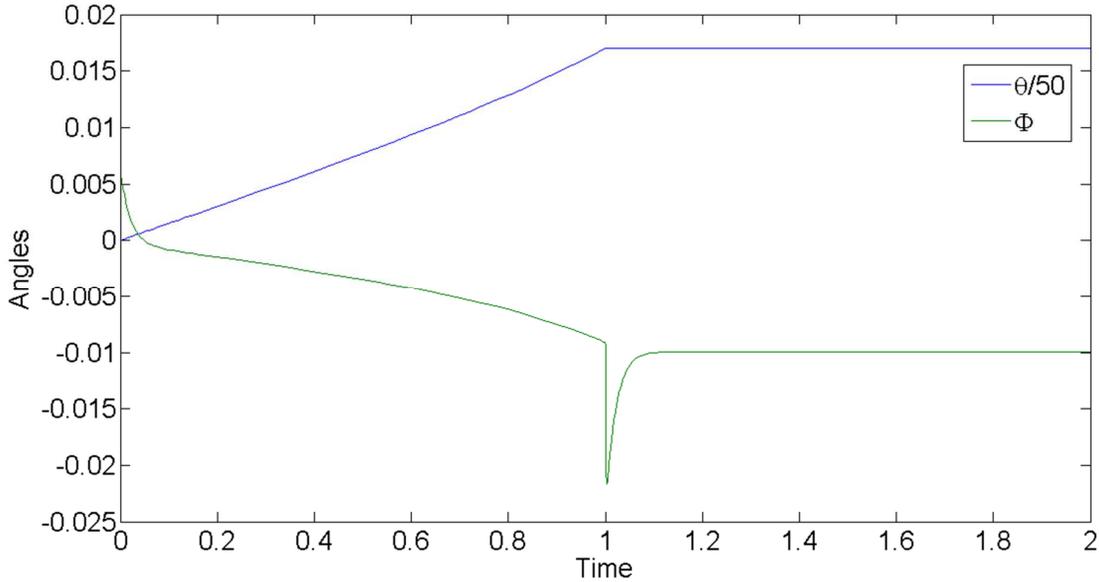

Figure 8 : *Entry into a turn, as per (59-60). The initial counter-steering is visible clearly. Note that* θ *is scaled down by 50 – its actual equilibrium value is about 45°.*

Clearly, Φ becomes briefly positive i.e. the bike turns to port side before crossing the zero and starting the turn to starboard. The lean angle becomes about 0.8 radians when the torque is withdrawn (in the plot I have scaled down θ by 50 to make them both of the same size), and Φ settles into the corresponding equilibrium value of −0.01 radians. Note that the equilibriation of θ is smooth with no overshoot or oscillation, as predicted from a heavily damped harmonic oscillator equation. The value of the torque here has been chosen as 15 units; if all units are understood to be SI then the torque estimate is a good approximation of its value in the actual situation, and the corresponding turn entry time of 1 second is also in good agreement with reality.

On the other hand, let us try the case where the driver attempts a turn without counter steering. In this situation he does not apply torque on the mobike, but just sets an initial desired steering angle and then watches the world go by. This is mathematically described by (53-4) with an initial condition Φ=0.01, θ=0 and $\dot{\theta} = 0$ (the desired turn this time is to port side). The response is below.

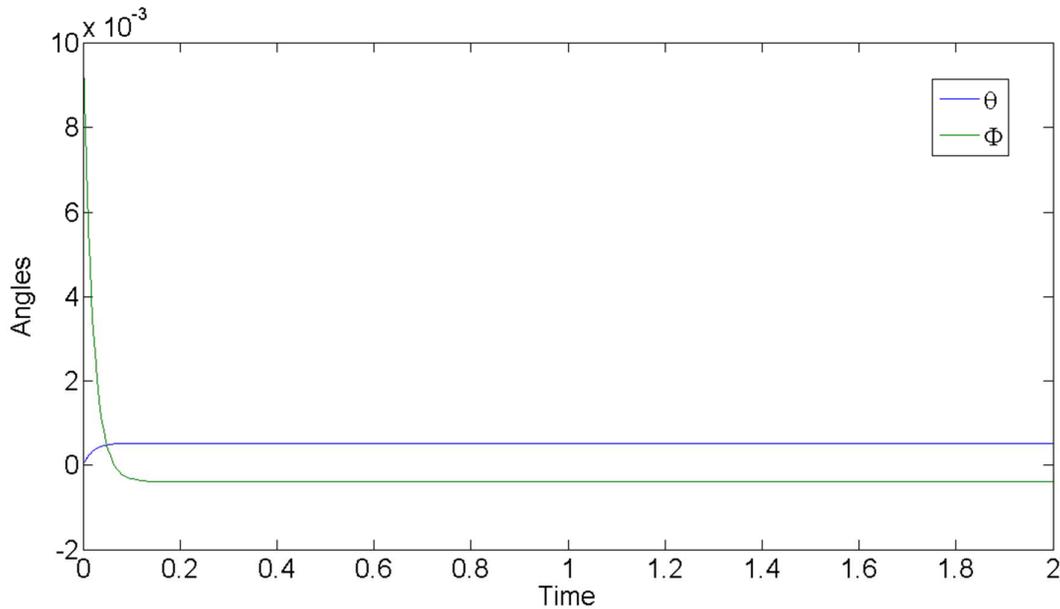

Figure 9 : *Attempted turn without counter-steering. The direction of the turn itself is wrong. Note also that this time there is no scaling on* θ.

Instead of the desired port side turn, the mobike settles into an extremely shallow turn to the starboard ! Thus our nonlinear motorcycle model clearly explains the phenomenon of counter-steering.

## Conclusion

I have already said a lot about the motorcycle and will now bring the Article to a rapid conclusion. Here, I have derived the nonlinear EOM of a motorcycle for the first time in Literature. The equations are (53-4) in the absence of an active driver and (59-60) in the presence of driver. These equations yield that the motorcycle is stable on straights and in turns. There is one zero eigenvalue because there is a turning radius for every lean angle and the motorcycle has no intrinsic preference for any radius. These equations can also explain quantitatively the phenomenon of counter-steering.

That said, there is considerable scope for refining the accuracy of this model. The effect of inclined steering axis can be taken into account, as well as that of the frame principal axes being at an angle to its geometric axes. The fastness approximation can also be relaxed and terms of higher order calculated, to improve the model accuracy at low speeds, and make it relevant for everyday operation of bicycles. All these modifications are however of an algebraic nature – they cannot impact the structure of the force and torque balance equations, and hence the feasibility of obtaining an explicit EOM. In this respect, the Newtonian method I have presented here is more versatile than the Lagrangian approach adopted in the literature. The source of this versatility is the ease of implementation of the non-holonomic constraints.

I hope that the conclusions of this Article will have applications in the design of motorcycles and in the strategizing of motorcycle races. An explicit analytical description of the transition from straight to turn can be used to optimize the performance of a racing motorcyclist in a turn. It can also be employed to adjust the various machine parameters so as to enable fastest entry into and exit from turn. At the same time, the analysis poses interesting theoretical questions – foremost among them being the origin of the almost unbelievable angle formula (56). The large role played by the apparently minuscule variable $\Phi$ is also noteworthy. Hence this Article creates considerable potential for further investigation of both theoretical and applicational aspects.

$$* \quad * \quad * \quad * \quad *$$